\documentclass[fleqn,usenatbib]{mnras}

\usepackage{newtxtext,newtxmath}
\usepackage{tabularx}
\usepackage{color, colortbl}
\usepackage{placeins}

\definecolor{LightOrange}{rgb}{1, 0.84, 0.6}
\definecolor{LightRed}{rgb}{1, 0.7, 0.7}
\definecolor{LightGreen}{rgb}{0.76, 0.94, 0.76}
\definecolor{LightYellow}{rgb}{0.99, 0.9, 0.4}

\usepackage[T1]{fontenc}
\usepackage{ae,aecompl}

\usepackage{graphicx}	
\usepackage{amsmath}	
\usepackage{hyperref}
\usepackage{upgreek}
\usepackage{gensymb}

\newcommand{\angstrom}{\text{\normalfont\AA}}

\newcommand{\magsqm}{${\mathrm{mag}_{\mathrm{SQM}}\,\mathrm{arcsec}^{-2}}$}

\newcommand{\fluxunit}{$\mathrm{erg}\ s^{-1}\ \mathrm{cm}^{-2}\ \angstrom^{-1}$}
\newcommand{\trenderror}{$\pm$1.5} 

\title[Impact of environment and atmosphere on light pollution]{Long-term trends of light pollution assessed from SQM measurements and an empirical atmospheric model\thanks{Generated using Copernicus Climate Change Service (C3S) Information and Atmosphere Monitoring Service (CAMS) Information, 2019.}}

\author[Puschnig, Wallner, Schwope, N\"aslund]{
Johannes Puschnig,$^{1}$\thanks{E-mail: johannes@jpuschnig.com}
Stefan Wallner,$^{2,3}$
Axel Schwope,$^{4}$
Magnus N\"aslund$^{5}$
\\
$^{1}$Universit\"at Bonn, Argelander-Institut f\"ur Astronomie, Auf dem H\"ugel 71, D-53121 Bonn, Germany\\
$^{2}$ICA, Slovak Academy of Sciences, Dubravska cesta 9, 84503 Bratislava, Slovak Republic\\
$^{3}$Universit\"at Wien, Institut f\"ur Astrophysik, T\"urkenschanzstra{\ss}e 17, A-1180 Wien, Austria\\
$^{4}$Leibniz-Institut f\"ur Astrophysik Potsdam (AIP), An der Sternwarte 16, 14482 Potsdam, Germany \\
$^{5}$Department of Astronomy, Stockholm University, AlbaNova University Centre, SE-10691 Stockholm, Sweden\\
}

\date{Accepted 2022 October 15. Received 2022 October 11; in original form 2022 August 31}

\pubyear{2022}

\begin{document}
\label{firstpage}
\pagerange{\pageref{firstpage}--\pageref{lastpage}}
\maketitle

\begin{abstract}
We present long-term (4-10 years) trends of light pollution observed at 26 locations,
covering rural, intermediate and urban sites, including the three major
European metropolitan areas of Stockholm, Berlin and Vienna. Our analysis is based on i) night sky brightness (NSB)
measurements obtained with Sky Quality Meters (SQMs) and ii) a rich set of atmospheric data products.
We describe the SQM data reduction routine in which we filter for moon- and clear-sky data
and correct for the SQM ``aging'' effect using an updated version of the twilight method of Puschnig et al. (2021).
Our clear-sky, aging-corrected data reveals short- and long-term (seasonal) variations due to atmospheric changes.
To assess long-term anthropogenic NSB trends, we establish an empirical atmospheric model via
multi-variate penalized linear regression.
Our modeling approach allows to quantitatively investigate the importance of different
atmospheric parameters, revealing that surface albedo and vegetation have by far the largest impact
on zenithal NSB. Additionally, the NSB is sensitive to black carbon and organic matter aerosols at urban and
rural sites respectively. Snow depth was found to be important for some sites, while the total
column of ozone leaves impact on some rural places.
The average increase in light pollution at our 11 rural sites is 1.7\% per year. At our nine urban sites
we measure an increase of 1.8\% per year and for the remaining six intermediate sites we find an
average increase of 3.7\% per year. These numbers correspond to doubling times of 41, 39 and 19 years.
We estimate that our method is capable of detecting trend slopes shallower/steeper than
\trenderror\% per year.
\end{abstract}




\section{Introduction}
During the last decade, an ever increasing number of studies found evidence
that artificial light at night (ALAN) leads to negative consequences
-- not only for astronomy -- but also
for ecosystems \citep[e.g.][]{Longcore2004,Perkin2011},
biodiversity \citep[e.g.][]{Hoelker2010},
animals \citep[e.g.][]{Eisenbeis2006,Perkin2013,Mathews2015,Owens2020,Parkinson2020} and
human beings \citep[e.g.][]{Chepesiuk2009,Haim2013,Cho2015,Garcia-Saenz2018,Khodasevich2020,Menendez-Velazquez2022}.
Therefore, monitoring night sky brightness (NSB) was soon recognized as being an inevitable effort
in order to keep track of light pollution.
Various organisations and individuals around the globe started
to continuously measure the NSB using different methods and devices \citep{Haenel2018}.
One of the probably most widely used devices is the so called Sky Quality Meter (SQM).
Operational SQM networks are found e.g. in Austria \citep{Posch2018,Puschnig2020},
Spain \citep{Zamorano2015,Bara2019b}, Italy \citep{Bertolo2019} and the Netherlands \citep{Schmidt2020}.
Furthermore, several individuals have mounted SQMs at various sites on Earth \citep[e.g.][]{Puschnig2014a,Puschnig2014b,Zeljko2018}. \cite{Kyba2015}
have compiled many of these heterogene datasets to study the change of light pollution on a global scale.

Early light pollution studies based on SQM data revealed the strong impact of clouds
\citep{Kyba2011,Puschnig2014a,Jechow2019,Sciezor2020a} on the NSB. And more recently, the impact of
several atmospheric parameters, in particular
aerosol optical depth (AOD) and particulate matter (PM) were observed to show correlations
with NSB \citep{Sciezor2014,Posch2018,Sciezor2020b,Kocifaj2021}.
Also, other seasonal parameters such as e.g. surface albedo or vegetation \citep{Wallner2019,Puschnig2020} were
previously discussed in the literature as potential factor causing changes in
NSB.

More recently, \cite{Bara2021} and \cite{Puschnig2021} reported on the degradation of
the SQM sensitivity (i.e. darkening) with time, when the devices are used outdoors.
Using reference measurements from unused/unexposed SQMs at the beginning and end of
a multi-year time series, \cite{Bara2021} find that readings from their SQMs, located in Galicia,
need to be corrected by approximately 0.06~\magsqm\~yr$^{-1}$. \cite{Puschnig2021} have
shown that twilight may serve as a source for calibration and they find that their readings
(Vienna, Berlin and Stockholm) need to be corrected by approximately 0.03--0.05~\magsqm\~yr$^{-1}$.
A combined view on these results suggests that the ``aging effect'' depends on solar
irradiance, that is a function of geographic latitude.

The assessment of long-term trends of light pollution at a given site is thus 
a complex problem that involves knowledge of atmospheric parameters
as well as knowledge/derivation of the instrumental darkening of the SQM (or similar devices) with time.
In this paper we aim to solve the problem via the combination of
SQM data with atmospheric parameters obtained through the
Copernicus Climate Change Service (C3S) and the Copernicus Atmosphere Monitoring Service (CAMS; \citealt{CAMS}).
We establish a simple empirical model to predict NSB variations
due to changing atmospheric conditions. As a result, the purely ALAN-driven long-term
change of NSB is assessed.

The paper is organised as follows. In Section \ref{sec:data} we give an overview of our
SQM sites and the atmospheric parameters that are used in this study.
In Section \ref{sec:methods} we describe the data reduction routine that we apply to
the SQM data to select for moonless, clear sky measurements. The section also contains
the description of the empirical atmospheric model to predict NSB variations.
The derived long-term trends for our sites are presented in Section \ref{sec:results}.
Finally, we discuss the results and summarize the paper in Sections \ref{sec:discussion}
and \ref{sec:summary}.

\section{Measurements and Data Products}\label{sec:data}

\begin{table*}
\centering
\caption[Basic information of the measurement sites]{Basic information about our SQM
measurement sites, categorized into urban, intermediate and rural ones.}
\label{tab:sqmsites}
\begin{tabular}{llllrrr}
Code & Name                      & Latitude N   & Longitude E      & Elevation {[}m{]}   & data     & operating\\
     &                           &              &                  & (above sea level)   & from-to  & time in yrs. \\
\hline\hline
\multicolumn{7}{c}{\textit{urban}} \\
\hline
STO & Stockholm (AlbaNova University Center)                  &  N 59 21 12  & E 18 3 28 & 30 & Dec. 2014 -- Dec. 2021 & 7.1 \\
IFA & Vienna (Insitute for Astronomy)                     &  N 48 13 54  & E 16 20 3 & 250 & Apr. 2012 -- Dec. 2021 & 9.75\\
BA1 & Potsdam -- Babelsberg         &  N 52 22 48  & E 13 6 22 & 90 & Jan. 2011 -- Feb. 2021 & 10.2\\ 
GRA & Graz -- Lustbuehel            &              &               & & Jul. 2014 -- Jun. 2020 & 6.0\\
LSM  & Linz, Schlossmuseum        & N 48 18 19 & E 14 16 58     & 287 & Aug. 2014 -- Dec. 2021 & 7.4\\
LGO  & Linz, G\"othestra{\ss}e    & N 48 18 19 & E 14 18 30     & 259 & Jan. 2014 -- Dec. 2021 & 8.0\\
STY  & Steyr                     & N 48 2 57  & E 14 26 32     & 307 & Aug. 2014 -- Dec. 2021  & 7.4\\
TRA  & Traun                     & N 48 14 8  & E 14 15 11     & 269 & Jan. 2015 -- Dec. 2021  & 7.0\\
WEL  & Wels, Rathaus              & N 48 9 23  & E 14 1 29      & 317 & Aug. 2014 -- Dec. 2021  & 7.4\\
\hline
\multicolumn{7}{c}{\textit{intermediate}} \\
\hline
BRA  & Braunau                   & N 48 15 40 & E 13 2 41      & 351 & Jan. 2016 -- Dec. 2021  & 6.0\\
GRI  & Grieskirchen              & N 48 14 4  & E 13 49 33     & 336 & Jan. 2016 -- Dec. 2021  & 6.0 \\
FRE  & Freistadt                 & N 48 30 33 & E 14 30 7      & 512 & Jan. 2016 -- Dec. 2021  & 6.0 \\
MAT  & Mattighofen               & N 48 5 50  & E 13 9 6       & 454 & Jan. 2016 -- Dec. 2021  & 6.0 \\
PAS  & Pasching                  & N 48 15 31 & E 14 12 36     & 292  & Jan. 2015 -- Dec. 2021   & 7.0 \\
VOE  & V\"ocklabruck               & N 48 0 21  & E 13 38 43     & 434 & Jan. 2016 -- Dec. 2021  & 6.0 \\
\hline
\multicolumn{7}{c}{\textit{rural}} \\
\hline
FOA  & Mittersch\"opfl           & N 15 55 24 & E 48 5 3 & 880 & Jan. 2013 -- Jul. 2019 & 6.6 \\
BOD  & Nationalpark Bodinggraben & N 47 47 31 & E 14 23 38     & 641 & Jul. 2016 -- Dec. 2021 & 5.5 \\
FEU  & Feuerkogel                & N 47 48 57 & E 13 43 15     & 1628 & Jan. 2016 -- Dec. 2021  & 6.0 \\
GIS  & Giselawarte               & N 48 23 3  & E 14 15 11     & 902  & Jan. 2016 -- Dec. 2019  & 4.0\\
GRU  & Gr\"unbach                  & N 48 31 50 & E 14 34 30     & 918 & Jan. 2016 -- Dec. 2021  & 6.0\\
KID  & Kirchschlag -- Davidschlag   & N 48 26 31 & E 14 16 26     & 813 & Jan. 2016 -- Dec. 2021  & 6.0\\
KRI  & Krippenstein              & N 47 31 23 & E 13 41 36     & 2067  & Jan. 2016 -- Dec. 2021  & 6.0\\
LOS  & Losenstein, Hohe Dirn      & N 47 54 22 & E 14 24 40     & 982  & Jan. 2016 -- Dec. 2021  & 6.0\\
MUN  & M\"unzkirchen               & N 48 28 45 & E 13 33 29     & 486 & Jan. 2016 -- Dec. 2021  & 6.0\\
ULI  & Ulrichsberg, Sch\"oneben     & N 48 42 20 & E 13 56 44     & 935 & Jan. 2016 -- Dec. 2021  & 6.0\\
ZOE  & Nationalpark Z\"obloden    & N 47 50 18 & E 14 26 28     & 899 & Jan. 2016 -- Dec. 2021 & 6.0       
\end{tabular}
\end{table*}

\subsection{Night Sky Brightness Measurements in Stockholm, Berlin, Vienna and 23 more sites in Austria}
This study is based on long-term zenithal NSB measurements obtained with Sky Quality Meters (SQMs) located at various sites (see Table \ref{tab:sqmsites}),
including metropolitan areas of Stockholm, Berlin and Vienna.
In previous studies, we already used parts of the data that is also included in this work, e.g.
\cite{Puschnig2014b} and \cite{Puschnig2014a} studied the influence of the Moon, clouds and other environmental effects on the night sky brightness
over Potsdam and Vienna using the first 1--2 years of SQM data.
A detailed description and quantification of the light pollution level at the Upper Austrian sites
is found in \cite{Posch2018}, who examined data obtained during the years of 2015 and 2016.

We cover very remote locations such as Krippenstein on the Dachstein
plateau ($\sim$2000\,m above sea level) as well as large metropolitan areas
such as Stockholm (STO), Vienna (IFA) or Potsdam-Babelsberg (BA1), located
$\sim$23\,km to the southwest of the center of Berlin.

The measurements are taken in an automated way, with LAN-attached
SQM devices (model SQM-LE) located in weather-proof housings.
The SQMs situated in Upper Austria are run by the provincial government of Upper Austria.
They take NSB measurements every minute. Most other SQMs (STO, IFA, FOA, GRA) provide a reading every 7 seconds which corresponds to a frequency of
0.143 Hz and BA1 even takes a measurement every $\sim$2 seconds.

\subsection{Atmospheric data products}
%
%
We make use of open access climate variables from ERA5 \citep{ERA5}, the fifth major global
reanalysis data produced by the European Centre for Medium-Range Weather Forecasts (ECMWF).
ERA5 is developed through the Copernicus Climate Change Service (C3S).
The data are based on a reanalysis of a large set of ground-, air- and satellite-based
measurements. Data are available with hourly validity time at a spatial resolution of
0.28$^{\degree}$ x 0.28$^{\degree}$ in latitude and longitude, corresponding to $\sim$30\,km x 30\,km).

Using the python package \texttt{cdsapi} provided by the
\textit{Atmosphere Data Store}\footnote{\url{https://ads.atmosphere.copernicus.eu}},
we downloaded for all our sites and dates more than 70
available atmospheric parameters.
From those we initially identified quantities that may
have impact on zenithal NSB measurements.
This pre-selection is mainly driven by our understanding of how the environment and
the atmosphere impact the NSB, i.e. we chose parameters that may alter the NSB via scattering of light (e.g. aerosols and particles)
as well as parameters that may enhance the fraction of upward light (e.g. albedo, snow cover) or enhance the fraction of
light that is reflected back to the ground (e.g. cloud cover).
An overview of the
selected parameters is found in Table \ref{tab:atmospheric_parameters}.

%
We note that aerosol optical depths and particulate matter are not available from ERA, but are
provided through CAMS \citep{CAMS}, the Copernicus Atmosphere Monitoring Service global reanalysis.
Daily Data products are available in steps of six hours. Due to a bug that caused AOD and PM to not
be computed for analysis time\footnote{\url{https://confluence.ecmwf.int/pages/viewpage.action?pageId=153393473}},
we had to use the 3-hour forecasts with a validity time of 3am UTC. 
CAMS native spatial resolution is 0.8$^{\degree}$ before 21 June 2016, and 0.4$^{\degree}$ henceforth,
corresponding to $\sim$80\,km x 80\,km and $\sim$40\,km x 40\,km.

\subsubsection{Description of selected atmospheric parameters}
An overview of our initial set of parameters used for statistical analysis is given in Table \ref{tab:atmospheric_parameters}.
In the following we briefly describe the physical underpinning of the most relevant parameters. 

\textit{Albedo (aluvd)} is defined as the fraction of incident radiation that is reflected by a surface.
Its numerical value thus ranges from 0 (no reflection) to 1 (all incident radiation is reflected).
It varies with the type of surface (e.g. its roughness) and wavelength. For example, the broad-band albedo of
grassland is a few percent only \citep{Briegleb1982,Briegleb1986,Coakley2003}, with twice as much reflection in the
near infrared than in the
visible
spectral range. On the other side, surfaces covered with snow may
reflect more then 90 percent of the incident radiation, with snow albedo being significantly higher in the
visible range than in the near infrared \citep{Roesch2002}.
In this study, we make use of the \textit{UV-optical albedo for diffuse radiation} (\textit{aluvd}),
that is measured within a range of 300 to 700nm and thus matches the SQM band very well.
It is thus expected that variations in surface albedo positively correlate with
the night sky brightness.

\textit{Leaf Area Index (lai)} is defined as the one-sided green leaf area per unit ground area
\citep{Boussetta2011}. It is thus a dimensionless number ($\mathsf{m^2\ m^{-2}}$).
In ERA5, the leaf area index is split into high (\textit{lai$_{hv}$}) and low vegetation (\textit{lai$_{lv}$}).
High vegetation consists of evergreen trees, deciduous trees, mixed forest/woodland, and interrupted forest and
low vegetation covers grass, shrubs as well as water and land mixtures. For our study we use the sum of both available quantities.
High vegetation impacts the NSB via blocking of upward light, while low vegetation has an impact on the
reflection of downward light. We expect a negative correlation between NSB and the leaf area index.

Both, the leaf area index and albedo are based on observations with the VEGETATION sensor on
board the SPOT satellite, that carries out measurements every 10 days using a composite
observation from a 30 days moving window at 1/112 degree spatial resolution, corresponding to
roughly 1 km at the equator \citep{Boussetta2014}.

\textit{Black Carbon} (\textit{bcaod550}) is a direct consequence of
anthropogenic activities, in particular black carbon is a result of incomplete combustion of
fossil fuels (38\%), biomass (42\%) and biofuels (20\%) \citep{Bond2004,Cao2014,Schwarz2006}.
Black carbon in the atmosphere is capable of absorbing and scattering of radiation. The atmospheric black carbon aerosol
optical depth measured at 550\,nm (\textit{bcaod550}) has thus the potential of having a large impact on our NSB measurements.
Note that black carbon may also have an impact on the surface albedo, e.g. when it is deposited on snow and ice.

\begin{table}
\centering
\caption[Overview of atmospheric parameters]{Overview of atmospheric parameters, i.e the features in the model}
\label{tab:atmospheric_parameters}
\begin{tabular}{ll}
\hline \hline
aluvd    & UV-optical albedo for diffuse radiation \\
lai      & leaf area index         \\
bcaod550  & black carbon aerosol optical depth     \\
duaod550 & dust aerosol optical depth              \\
ssaod550 & sea salt aerosol optical depth          \\
omaod550 & organic matter aerosol optical depth    \\
pm10     & particulate matter 10$\mu$m                   \\
pm2p5    & particulate matter 2.5$\mu$m                 \\
pm1    & particulate matter 1$\mu$m\\
tcwv   & total column water vapour \\
tcw    & total column water \\
wind   & wind 10m \\
tco3   & total column ozone \\
tcc     & total cloud cover \\
sd   & snow depth \\
\end{tabular}
\end{table}

\section{Methods}\label{sec:methods}

\subsection{Atmospheric model}\label{sec:atmospheric_model}
We aim to find a set of atmospheric parameters that most directly
impact (zenithal) NSB measurements.
In particular, we attempt to separate
fundamental correlations between the NSB and the atmosphere from those that are
indirect consequences of covariance among atmospheric metrics.
We distinguish these underlying relations through variable selection:
For each NSB data (as a target variable), we compose an empirical predictive model
using a set of atmospheric parameters (feature variables) that carry most predictive power.
This is an effective way to collapse a high-dimensional data set into a concise model.

The basis of this analysis is a multi-variable penalized linear regression.
That is, we restrict the model functional forms to simple linear combinations
of variables (including an intercept term).
The regression is done independently for each NSB measurement/chunk ($\Delta_\mathrm{sky-obs}$) as a target variable, using all
the atmospheric parameters plus time (to account for temporal trends) as available features.
With this regression setup, we perform a \textit{lasso} model
fit \citep{lasso} and use the Bayesian Information
Criterion (BIC; \citealt{Schwarz1987}) for automated feature/model selection. This
is implemented with the \texttt{LassoLarsIC} function in the
\texttt{scikit-learn} Python package. In detail, for a linear
predictive model with the form $\hat{y}_i = \beta_0 + \sum_{j=1}^m \beta_j x_{ij}$,
the lasso regression minimizes the following function:
\begin{equation}
    \frac{1}{2n} \sum_{i=1}^n\ (y_i - \hat{y}_i)^2\ + \alpha\ \sum_{j=1}^m\ |\beta_j|.
\end{equation}
The indices i and j run through the NSB data and the atmospheric features respectively.
The parameter $\alpha$ is a hyper-parameter, so that the second term in
the equation adds a penalty for the use of any non-zero slope in the fitted model.
This particular ``regularization'' term is the reason that the lasso as a regression
method can also be used for variable selection. The lasso regression yields a
best-fit model that minimizes the equation for each choice of the $\alpha$ parameter.

In practice, we produce 16 different predictive atmospheric models.
Starting with a model based on all 15 atmospheric and ground features plus time,
we expect to find the best agreement between observed NSB variations and the models.
However, we also aim to quantify the impact of each parameter. Hence, we iteratively
remove input features and re-calculate the models. This will allow us to use the residual after subtracting
the model from the data as a measure of the importance of individual input features
for the NSB modeling.

Note that due to the ``regularization'' term not all input features may survive
the lasso method and only those parameters
that have most direct impact on the NSB will be included in the final model.

\begin{figure*}
\centering
        \includegraphics[width=1.0\columnwidth]{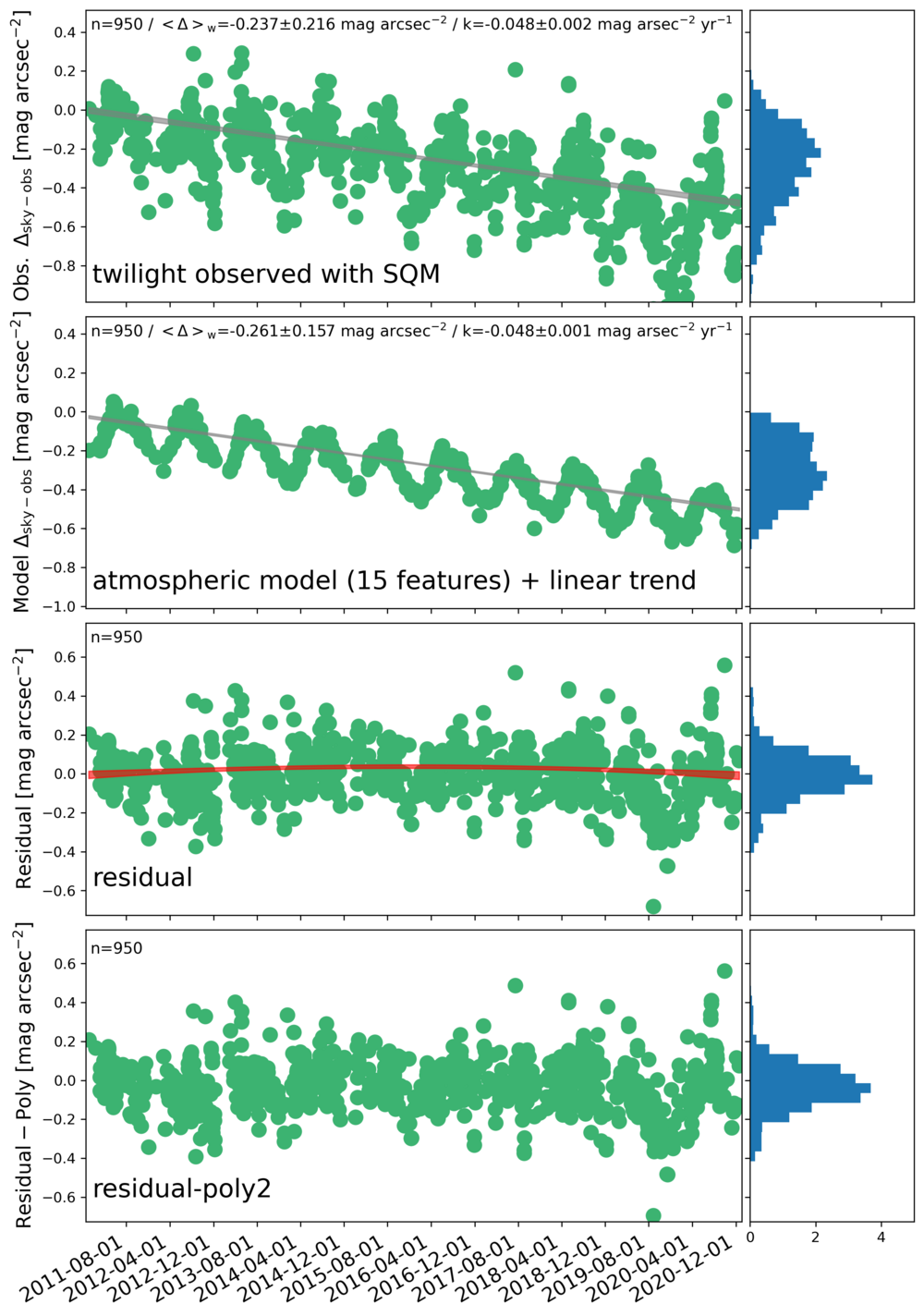}
        \includegraphics[width=1.0\columnwidth]{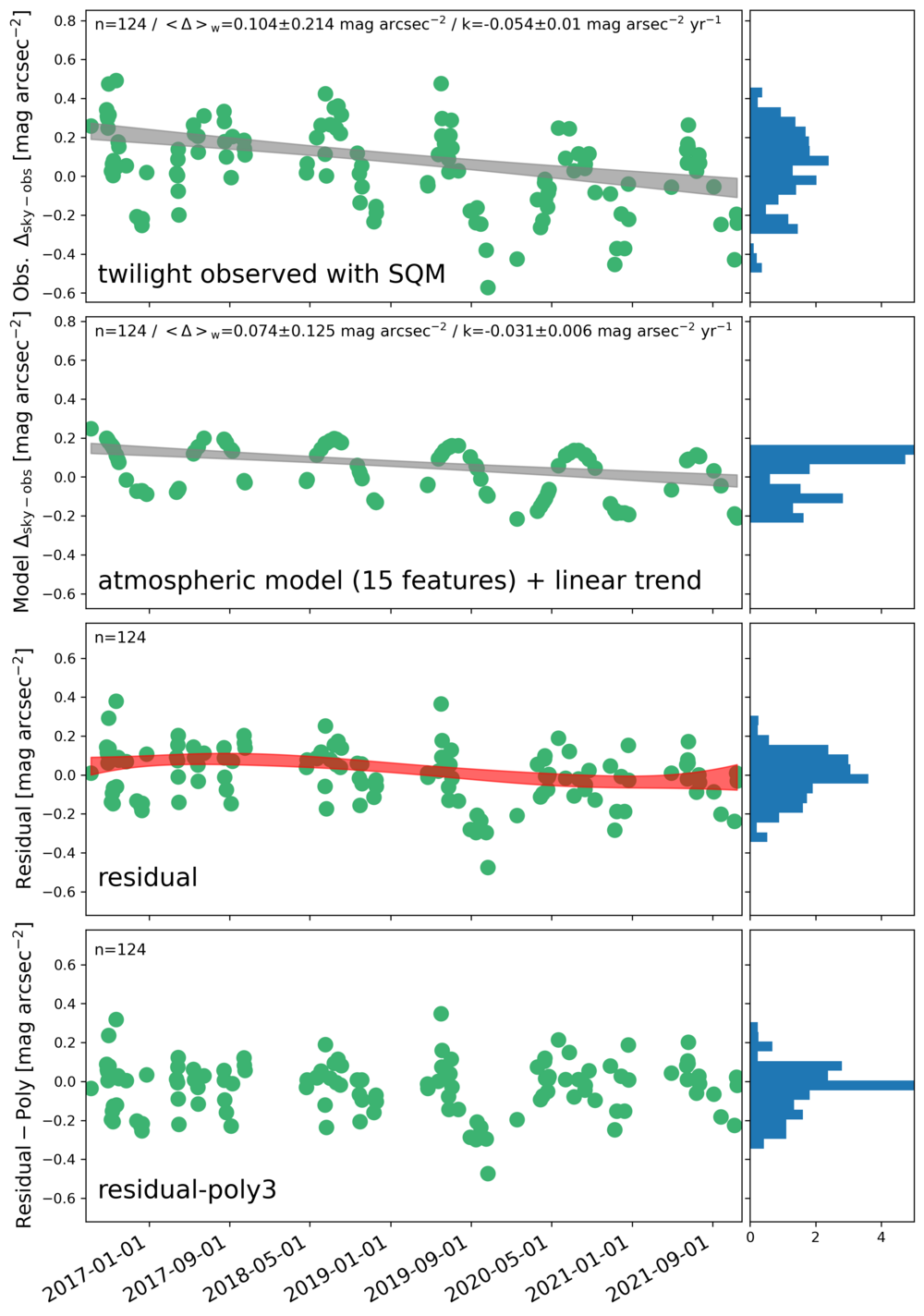}
        \caption[Aging effect from twilight method]{Measuring the SQM aging effect from twilight observations. \textit{Left:} Germany, Potsdam-Babelsberg (BA1; 10.2 years) \textit{Right:} Austria, Bodinggraben (BOD; 5.5 years).
        \textit{Topmost panels:} Patat model minus observed clear-sky SQM twilight data on y-axis versus time on x-axis. A strong degradation with time is revealed. Each point is a 5-minute-mean (BA1) or 10-minute-mean (BOD) brightness.
        \textit{2nd panels from top:} Prediction from empirical atmospheric model plus linear time trend.
        \textit{3rd panels from top:} The residual after subtraction of the top-panel data from the model in the 2nd panel. The red line and area show a polynomial fit and its
        standard deviation respectively.
        \textit{Bottom panels:} Final scatter after subtraction of the atmospheric model and the linear and polynomial trends.}
        \label{fig:BA1_BOD_aging}
\end{figure*}

\subsection{Correcting SQM readings for degrading sensitivity with time using the twilight method}
We apply an updated version of the ``twilight'' calibration as described in \cite{Puschnig2021} in order
to correct our multiple-year lasting SQM measurements for potential temporal changes in detector sensitivity.
This is the ``aging effect'', which is probably
caused by changing transmission of the 
SQM housing window due to UV light exposure over several years.
In brief, we utilize SQM observations of the zenithal NSB and compare the measurements
to the twilight model of \cite{Patat2006}. Moon and Sun altitudes
are calculated using the Python \texttt{ephem} package,
which provides an accuracy of approximately 1 arcsec.
Note that one could also use an inter-comparison of SQM
data obtained during twilight when the sun was at the same altitude. Assuming that the twilight
sky brightness is not affected by ALAN and remains constant, one can reveal any underlying
temporal sensitivity change of the measurement system.
In addition to this procedure, we now also account for atmospheric changes that
may impact the twilight observations. Thus, we utilize the same atmospheric model
as described in Section \ref{sec:atmospheric_model}. Any linear temporal trend obtained from the
modeling procedure is then attributed to the ``aging effect''.

Since we recognized at some stations (e.g. IFA) remaining non-linear trends,
we additionally fit polynomials up to 3rd order to the residuals. If the subtraction of
a polynomial further reduces the scatter (by more than 0.5 percent), we proceed and
use also the polynomial as a correction function, accounting for non-linear effects.

\begin{figure}
\centering
        \includegraphics[width=1.0\columnwidth]{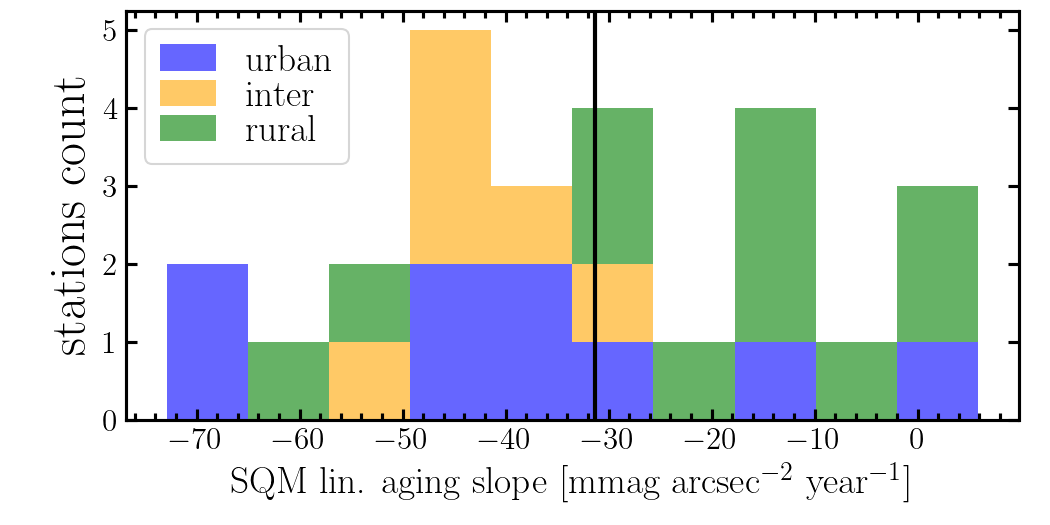}
        \caption[Histogram of linear aging slopes]{Histogram of the linear part of the aging slopes derived using the ``twilight'' method for our SQM sites.
        The vertical black line denotes the mean slope. Note that the unit on the x-axis is m\magsqm.}
        \label{fig:sqm_aging}
\end{figure}

A quantification of the degradation of SQM sensitivity with time is seen in Figure
\ref{fig:BA1_BOD_aging} for Potsdam-Babelsberg near Berlin and a rural
site in Austria (Bodinggraben).
Note the strong seasonal variation of NSB during twilight and how accurate the atmospheric model is able to predict them
for both the urban site and the rural site.
The histogram in Figure \ref{fig:sqm_aging} shows the distribution
of linear aging slopes found for all our 26 SQM stations.
We find a relatively large variation between individual SQM sites with aging slopes
ranging from zero to $\sim$-0.075\,\magsqm~yr$^{-1}$. On average, the aging effect leads
to a darkening of $\sim$-0.031$\pm$0.020\,\magsqm~yr$^{-1}$.

\subsection{Extraction of moonless, clear sky SQM measurements}
We are interested in the anthropogenic contribution to the NSB. Thus,
we compare our SQM measurements to a modeled zenithal NSB calculated
from sky spectra available through the SKYCALC sky model \citep{Noll2012,Jones2013}.
The whole procedure of how we extracted SQM magnitudes from the model spectra is described in the 
Appendix \ref{sec:appendix_skymodel}. From this point on, we proceed
with the difference between the sky model (\textit{sky}) and the observed
NSB (\textit{obs}): ($\Delta_\mathrm{sky-obs}$).
Note that -- given our model constraints -- the SKYCALC sky models lead to yearly peak and valley
zenithal NSBs (due to changing starlight+zodiacal light) of 21.63 and 21.87\,\magsqm\ for our sites.
Comparing our yearly peak-to-valley difference of 0.24\,\magsqm\ to the ``GAIA map of the brightness of the natural sky'' \citep{Masana2022},
shows that this is a relatively low value. Figure 3 of \cite{Masana2022} suggests
a maximum yearly V-band variations of approx. 0.6\,mag\,arcsec$^{-2}$.
This discrepancy arises due to several facts. First, we evalute the SKYCALC models using (for the sake of simplicity)
constraints such as a fixed precipitable water vapor value of 5\,mm rather than a time-dependet value.
This has impact on the scattering of light in the atmosphere and will lead to lower NSB variations over the year.
Second, our sites are located at geographic latitudes between 50 and 60 degrees. This is higher than the example
shown in Figure 3 of Masana et al. Thus the contribution of zodiacal light to the zenithal NSB becomes lower at our
sites \citep[see Figure 10]{Masana2020}, further reducing the overall yearly zenithal NSB variation.

Our data reduction routine starts with splitting the SQM data into chunks of 45 minute length.
We have chosen this time span to be short enough to allow for multiple sampling points each night
(even during summer, except for Stockholm) and to be long/large enough to avoid stochasticity.
For each of these time chunks we calculate the mean NSB and the standard deviation. Also a
linear fit is performed and the maximum deviation from the fit line is determined.
For the mean time of each chunk we further lookup the moon altitude.
That way, we are prepared to downselect for observations obtained
during moonless and clear sky conditions. We only keep chunks that fulfill the
following constraints: i) the Moon is below the horizon, ii) the maximum deviation from the
linear fit line of any single measurement is lower than 0.04 \magsqm, iii) the standard deviation
of the measurements within the chunk is less than 0.02 \magsqm\ for SQMs with
a high sampling rate (BA1, IFA, STO, FOA) and 0.06 \magsqm\ for SQMs with a sampling
frequency of only 1/minute, iv) the slope of the linear fit is shallower than 0.13 \magsqm\ 
per hour.

This procedure basically reduces the SQM data to chunks of 45 minute length within which the NSB remains almost constant.
This is the case only when no clouds are present in zenith. Note that the capability of using the SQM
as cloud detector was previously demonstrated by \cite{Cavazzani2020}.
Finally, we lookup meteorological parameters for each of the data chunks and reject
data chunks where i) the large-scale total cloud cover was larger than 50 percent and 
ii) the snow depth was larger than 5\,cm (avoiding SQM readings when the sensor was potentially covered with snow).

\begin{figure*}
\centering
        \includegraphics[width=1\columnwidth]{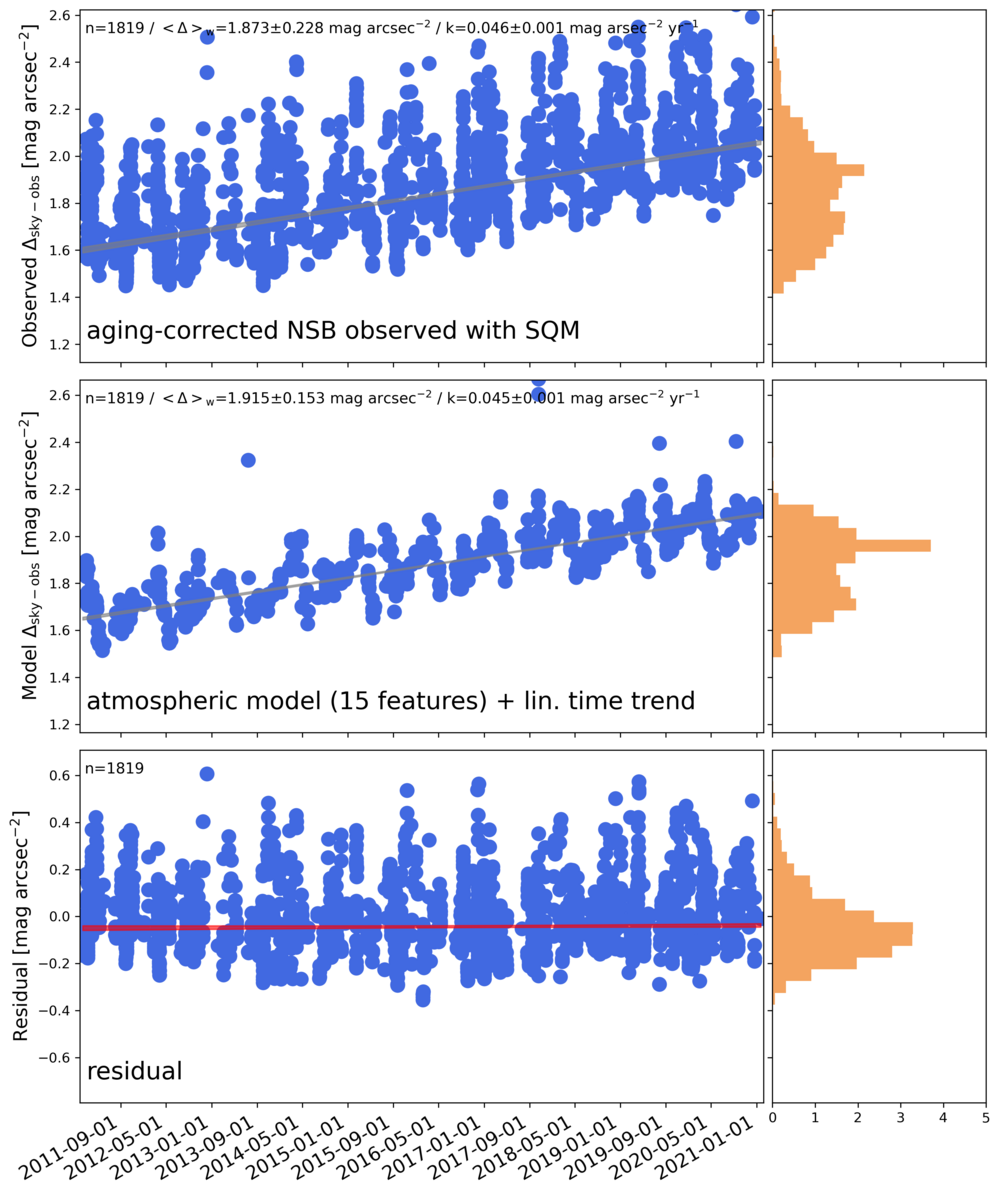}
        \includegraphics[width=1\columnwidth]{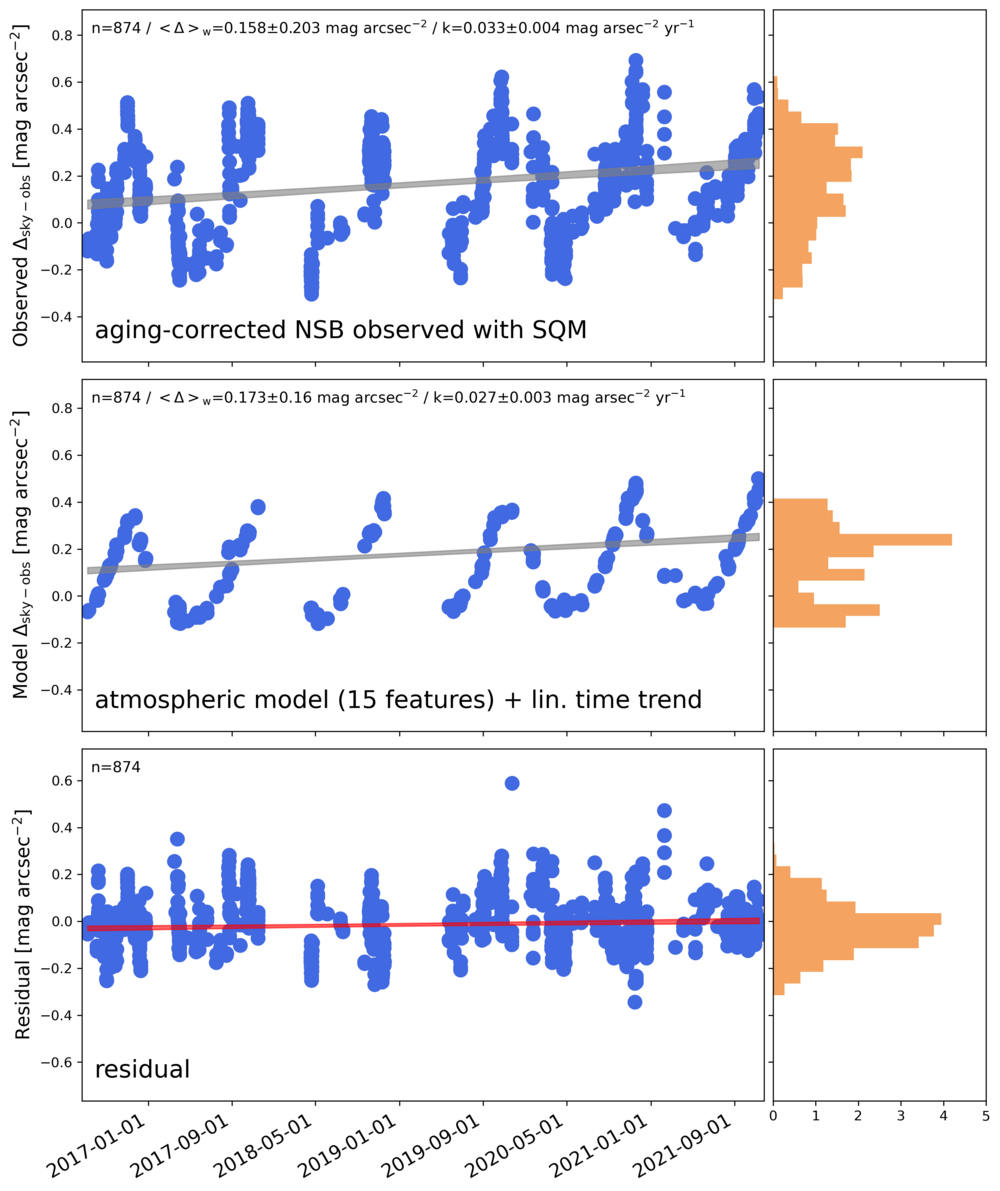}
        \caption[Long-term Trendanalysis]{\textit{Left:} Germany, Potsdam-Babelsberg (BA1). \textit{Right:} Austria, Bodinggraben (BOD).
        \textit{Top panels:} The difference between the predicted (skycalc model) and the observed clear-sky night sky
        brightness data within the SQM sensitivity range as a function of time. Each point represents the mean NSB within time chunks of 45 minute length.
        The gray area shows the $\pm$1.5$\upsigma$ range of a linear least square fit to the points.
        \textit{Mid panels:} The best-fit (penalized linear regression) model based on 15 atmospheric parameters obtained via the Copernicus Climate Change Service (C3S) Data Store.
        \textit{Bottom panels:} The residual after subtraction (model minus data).
        The panels to the right of the time series are histograms of the data points, using a binwidth of 0.05\magsqm.
        Note the observed seasonal variations of NSB for the rural station in Austria and how accurate the atmospheric model is able to predict them.}
        \label{fig:BA1_BOD_trend}
\end{figure*}


\subsection{Error estimation}
In order to estimate the error of our final trend slopes expressed in \magsqm\ yr$^{-1}$
we take into account uncertainties due to the aging correction and the trend fitting.
Both errors are estimated through examination of linear fits through the residuals,
e.g. bottom panels in Figures \ref{fig:BA1_BOD_aging} and \ref{fig:BA1_BOD_trend}.
We calculate the linear regression uncertainties from the mean squared errors which
are the trace (equal to the sum of elements on main diagonal) of the error covariance matrix.
That way, we find 1-sigma uncertainties due to the aging correction ranging from 0.001 to 0.006\,\magsqm\ yr$^{-1}$
with a mean uncertainty of 0.0035\,\magsqm\ yr$^{-1}$.
The 1-sigma uncertainties
resulting from the linear trend analysis
vary between 0.001 and 0.008\,\magsqm\ yr$^{-1}$
with a mean uncertainty of 0.0045\,\magsqm\ yr$^{-1}$.
Using a conservative approach, we finally calculate the \textit{absolute maximum 1-sigma error} for our routine
simply via addition: 0.0035\,+\,0.0045\,=\,0.0080\,\magsqm\ yr$^{-1}$.
We thus conclude that our method is capable of detecting trends that are shallower/steeper than
$\pm$0.016\,\magsqm\ yr$^{-1}$, corresponding to the $\pm$2-sigma level.
Expressed on a linear scale, our trend detection limit is thus \trenderror\ percent per year.

\section{Results}\label{sec:results}

\subsection{Long-term trends}
After selecting aging-corrected SQM data chunks as explained in Section \ref{sec:methods},
one can easily see in the top panel of Figure \ref{fig:BA1_BOD_trend} that even under clear and moonless 
conditions individual SQM measurements are prone to large variations with peak-to-valley
differences up to $\sim$1\magsqm, corresponding to a factor of 2.5. At rural sites
such as e.g. BOD, the scatter is dominated by a seasonal variation, caused by
variations of albedo (e.g. enhancing NSB during winters due to snow cover) and vegetation
(darkening during summers due to blocking of light on leafs).
In urban areas (e.g. BA1) such seasonal effects are typically weaker and other parameters seem to gain importance,
e.g. atmospheric aerosols or particulate matter.

Pronounced variations and the scattering of data points hamper the assessment of long-term trends.
For example, non-uniform, stochastic sampling of a periodic function (seasonal variations) may introduce a bias that 
leads to a spurious trend when performing a linear regression. For that reason, we aim to
model and remove the impact of the atmosphere. To do so, we use atmospheric products
freely available from the Copernicus Earth observation program. Our empirical atmospheric
model is then found from a penalized linear regression method (lasso) as explained in Section \ref{sec:methods}.

The middle panel in Figure \ref{fig:BA1_BOD_trend} makes evident that our model is capable of predicting
the bulk of the seasonal NSB variations solely from the atmospheric data. The long-term linear trends we find
for Potsdam-Babelsberg (BA1) and Bodinggraben (BOD) are 45$\pm$16 and 27$\pm$16\,m\magsqm\ yr$^{-1}$, respectively.
This corresponds to an increase of $\sim$4$\pm$1.5\% and 2$\pm$1.5\% per year.

The average increase in light pollution at our 11 rural sites is 1.7$\pm$1.5\% per year. At our nine urban sites
we measure an increase of 1.8$\pm$1.5\% per year and for the remaining six intermediate sites we find an
average increase of 3.7$\pm$1.5\% per year. These numbers correspond to doubling times of 41, 39 and 19 years.
Results for all 26 stations are found in Table \ref{tab:long_term}.

\begin{table*}
\centering
\caption[Long-term trends]{For each SQM station in \textit{column 1}, the linear slope of the aging effect given in m\magsqm\ year$^{-1}$ is shown in \textit{column 2}.
\textit{Column 3} contains the order of the polynomial used to account for non-linear sensitivity changes of the SQM.
The long-term linear trend slope in m\magsqm\ year$^{-1}$ is found in \textit{column 4}. All features that survived the lasso regression are listed in \textit{column 5}
and the total number of data chunks are given in \textit{column 6}.
The \textit{last column} contains the Pearson R correlation coefficient for the full model (using 16 features). The best fits with R>0.8 are BOD, FEU, KRI, STO.}
\label{tab:long_term}
\begin{tabular}{lrrrlrr}
Code & aging & polyn.   & trend & features that    & data     & R \\
     & slope & order    & slope & survived lasso   & count    & \\
\hline \hline
\multicolumn{7}{c}{\textit{urban}} \\
\hline
\rowcolor{LightOrange}
BA1 & -48 & 2 & 45 & time, lai, bcaod550, duaod550, omaod550, ssaod550 & 1819 & 0.74 \\
\rowcolor{LightRed}
GRA & -73 & 3 & 49 & time, aluvd, lai, bcaod550, pm1, duaod550, omaod550, ssaod550, pm2p5, tcw, sd, wind10, tco3, tcc & 846 & 0.58 \\
\rowcolor{LightOrange}
IFA & -68 & 2 & 35 & time, aluvd, lai, bcaod550, pm1, duaod550, omaod550, pm10, pm2p5, tcwv, wind10, tco3, tcc & 1506 & 0.59 \\
\rowcolor{LightGreen}
LGO & 6 & 3 & -12 & aluvd, lai, pm1, duaod550 & 315 & 0.59 \\
\rowcolor{LightGreen}
LSM & -35 & 3 & -16 & aluvd, lai, sd & 424 & 0.6 \\
\rowcolor{LightYellow}
STY & -15 & 2 & 19 & time, aluvd, lai, bcaod550, duaod550, omaod550, ssaod550, pm2p5, sd, wind10, tco3 & 1335 & 0.51 \\
\rowcolor{LightYellow}
TRA & -37 & 3 & 29 & time, aluvd, lai, duaod550, omaod550, ssaod550, wind10, tco3 & 410 & 0.53 \\
\rowcolor{LightGreen}
WEL & -29 & 3 & -9 & time, aluvd, lai, pm1, duaod550, omaod550, ssaod550, wind10 & 1101 & 0.64 \\
\rowcolor{LightOrange}
STO & -43 & 0 & 37 & time, aluvd, lai, duaod550, omaod550, tcwv, sd & 913 & 0.82 \\
\hline
\multicolumn{7}{c}{\textit{intermediate}} \\
\hline
\rowcolor{LightYellow}
BRA & -45 & 3 & 31 & time, aluvd, lai, pm1, duaod550, omaod550, ssaod550, tco3 & 970 & 0.58 \\
\rowcolor{LightYellow}
FRE & -36 & 0 & 21 & time, aluvd, lai, bcaod550, duaod550, omaod550, ssaod550, sd, tco3 & 960 & 0.61 \\
\rowcolor{LightOrange}
GRI & -45 & 0 & 46 & time, aluvd, lai, bcaod550, pm1, ssaod550, pm10, pm2p5, tcwv, sd, wind10, tco3 & 893 & 0.67 \\
\rowcolor{LightRed}
MAT & -48 & 3 & 90 & time, aluvd, lai, bcaod550, duaod550, ssaod550, tcwv, sd, wind10, tco3, tcc & 979 & 0.64 \\
\rowcolor{LightGreen}
PAS & -29 & 3 & 7 & time, aluvd, lai, bcaod550, pm1, duaod550, omaod550, pm10, pm2p5, tcwv, sd, wind10, tco3, tcc & 564 & 0.6 \\
\rowcolor{LightRed}
VOE & -50 & 0 & 50 & time, aluvd, lai, bcaod550, duaod550, omaod550, ssaod550, pm10, tcwv, sd, wind10, tco3, tcc & 678 & 0.54 \\
\hline
\multicolumn{7}{c}{\textit{rural}} \\
\hline
\rowcolor{LightGreen}
FOA & -28 & 3 & 7 & time, aluvd, lai, bcaod550, duaod550, omaod550, ssaod550, pm10, tcw, sd, wind10, tco3, tcc & 1122 & 0.65 \\
\rowcolor{LightYellow}
FEU & -5 & 3 & 21 & aluvd, tco3 & 848 & 0.81 \\
\rowcolor{LightRed}
GIS & 0 & 2 & 62 & time, aluvd, lai, bcaod550, duaod550, pm10, wind10, tco3, tcc & 343 & 0.59 \\
\rowcolor{LightGreen}
GRU & -17 & 0 & 11 & aluvd, lai, bcaod550, duaod550, omaod550, tcwv, sd, wind10, tco3, tcc & 1265 & 0.63 \\
\rowcolor{LightGreen}
KID & -11 & 0 & -11 & time, aluvd, lai, bcaod550, duaod550, omaod550, tcwv, sd, tco3, tcc & 1069 & 0.55 \\
\rowcolor{LightYellow}
KRI & -24 & 3 & 22 & time, aluvd, lai, bcaod550, tco3 & 800 & 0.82 \\
\rowcolor{LightGreen}
LOS & -51 & 0 & 1 & aluvd, lai, omaod550, ssaod550, tcwv, tco3 & 913 & 0.54 \\
\rowcolor{LightRed}
MUN & -58 & 3 & 61 & time, aluvd, lai, bcaod550, duaod550, sd, tco3 & 860 & 0.72 \\
\rowcolor{LightYellow}
BOD & -31 & 3 & 27 & time, aluvd, lai, bcaod550, omaod550, ssaod550, tco3 & 874 & 0.84 \\
\rowcolor{LightGreen}
ZOE & -17 & 3 & 5 & aluvd, lai, duaod550, omaod550, tco3 & 817 & 0.78 \\
\rowcolor{LightGreen}
ULI & -1 & 3 & 3 & time, aluvd, lai, bcaod550, duaod550, omaod550, pm10, tcw, sd, wind10, tco3, tcc & 908 & 0.68 \\
\end{tabular}
\end{table*}

\subsection{Impact of the atmosphere}
As explained in Section \ref{sec:methods}, we perform a multi-variate penalized linear regression to
identify the importance of individual atmospheric parameters.
In detail, we calculate models with increasing complexity, i.e. with an increasing number of atmospheric input features.
Starting with a model based on only a single variable (time), any potential long-term linear trend is revealed.
Then, we consecutively add single atmospheric features (those listed in Table \ref{tab:atmospheric_parameters})
to the model in order to study the relative importance of each parameter. That is,
we compare the scatter (standard deviation) of the residuals after subtracting the model from the data.

\begin{figure}
\centering
        \includegraphics[width=1.0\columnwidth]{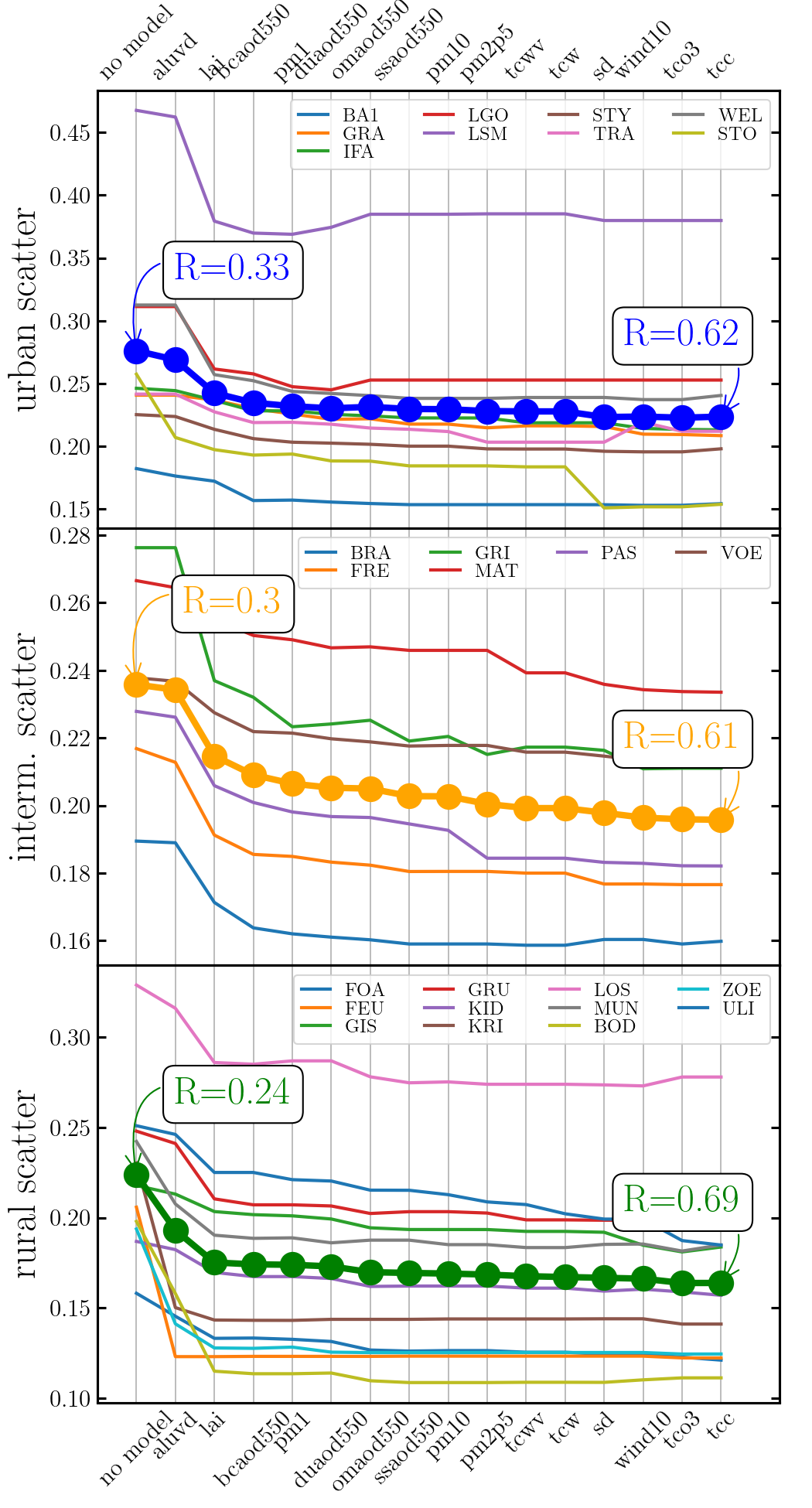}
        \caption[Model Comparison]{Relative impact of atmospheric parameters. The number of model input features increases from left to right along the x-axis, starting
        with a solely linear time dependence (very left; labelled as no model) to 16 features for the full model (right end; time plus 15 atmospheric features). The y-axis
        is the residual scatter (standard deviation) found after subtracting the model from the data. The thick curves with points indicate the mean scatter over all stations in the panel.
        The numbers in the boxes are the Pearson correlation coefficients (R) without any atmospheric model (left) and with the full model (right).}
        \label{fig:model_comparison}
\end{figure}

As shown in Figure \ref{fig:model_comparison}, additional features reduce the residual scatter. In particular, the Figure
shows that at all our sites, the leaf area index (\textit{lai}) and surface albedo (\textit{aluvd}) are the most important model input parameters
that reduce the final scatter by a large fraction.
Snow depth (\textit{sd}) in principle may firmly enhance the NSB. However, we did not expect that \textit{sd} has a strong impact
on our models, since for the bulk of our locations \textit{sd} is typically zero most of the time. Moreover, we rejected data points
when the snow cover was more than 5\,cm (to avoid observations when the SQM sensor was covered with snow).
Yet, we do see a decreasing residual scatter due to \textit{sd}
for some rural sites (BOD, FEU, KRI, ZOE) and we observe that snow depth even has a strong effect in Stockholm (STO).
Variations of albedo mostly impact our rural sites, while urban and intermediate sites are less affected.
In contrast to that, the leaf area index (\textit{lai}) helps in reducing the scatter of all sites
by a large degree, be it rural, intermediate or urban.

Atmospheric black carbon (\textit{bcaod550}) is an important ingredient for our models of urban and (in particular) intermediate sites,
while it has almost no impact on rural sites.
Interestingly, our coarsly spatially resolved atmospheric data, does not suggest any strong correlation between NSB
and large particulate matter (\textit{pm10}). However, for smaller grains, we do see that \textit{pm2p5} and \textit{pm1}
reduce the residual scatter of the bulk of our urban and intermediate sites.
Organic matter in the atmosphere (\textit{omaod550}) plays mostly only a minor role at all rural sites, but at one station (GIS),
\textit{omaod550} seems to be important for the modeling.

Ozone makes no difference for the modeling of our urban and intermediate sites. However, models that include ozone as a feature reduce the residual scatter
of some rural sites such as ULI, GIS, GRU.
Finally, we find that the 10-meter wind speed (\textit{wind10})
improves the match between the models and NSB observations for some of our sites
(GRA, GRI, GIS).
The correlation between wind speed and NSB is a non-causal secondary effect, that may be the result of an increase in aerosol transport 
when wind speed increases. We reckon that due to the coarser temporal resolution of our AOD data (one measurement per six hours) compared to the wind
data (hourly data), at sites where aerosol abundance correlates with wind speed, the wind parameter may serve as a proxy for AOD variations
at shorter time scales.

For each of our models and sites we calculate the Pearson correlation coefficient $R$ (see Table \ref{tab:long_term}). While the correlation between the SQM data and a solely linear
trend is weak ($R\sim0.3$ on average), the correlation between the full atmospheric model and the SQM observations becomes strong ($R\sim0.7$ on average).
For some of the sites (BOD, FEU, KRI, STO) the Pearson correlation coefficient even increases to values greater than 0.8, indicative for a very strong
correlation.

\section{Discussion}\label{sec:discussion}
Our long-term analysis of zenithal NSB measurements reveal that it is not the urban regions
that show the largest increase in light pollution, but rather the intermediate regions.
This may be explained by the fact that the installation of additional lighting points
in urban areas contributes relatively less to the NSB, as in urban areas the overall brightness level
is already extremely high. On the other side, when intermediate areas develop,
new or upgraded lights have a much larger impact on the NSB, as it is relatively darker
than in urban areas. More so, in rural regions. There, even only few additional lights may
have a recognizable impact.
Having said that, Stockholm is probably an exception of that average rule, because it is our
brightest site and still shows a very strong increase.

\begin{figure}
\centering
        \includegraphics[width=1.0\columnwidth]{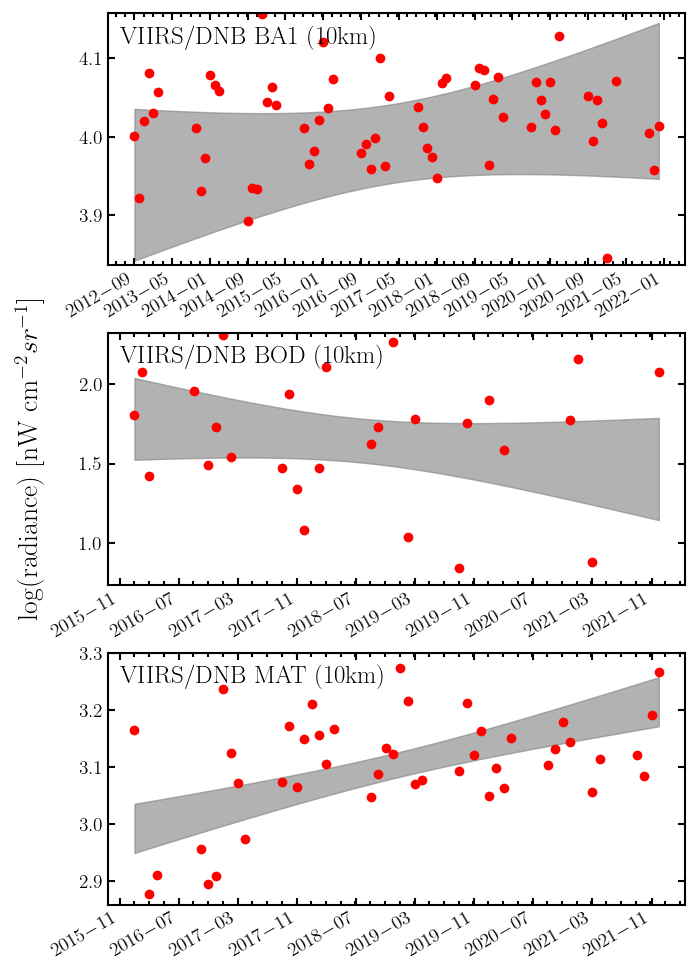}
        \caption[VIIRS trends]{Radiance trends for three of our sites
        obtained through the ``Radiance Light Trends'' web application.
        A radius of 10km around our SQM sites was chosen for summing up radiances. Each red point
        is a monthly average and the gray shaded area represents the uncertainty of a linear fit.}
        \label{fig:viirs_trends}
\end{figure}

For three of our stations (BA1, BOD, MAT) we utilize the
``Radiance Light Trends'' web application\footnote{\url{https://lighttrends.lightpollutionmap.info}},
which provides monthly mean radiance measurements obtained with the ``Day-Night Band'' (DNB) of the
``Visible Infrared Imaging Radiometer Suite'' (VIIRS) aboard the
Suomi-NPP
satellite. The resulting long-term radiance trends are shown in Figure \ref{fig:viirs_trends}.
The VIIRS monthly averages show a large scatter and no seasonal variation. This is caused by several
factors. First of all, VIIRS data is only available for our sites between
September and March, because only during these months the satellite passes during astronomical night.
This may drastically reduce the dynamic range of the observations and basically filters out
seasonal trends. Thus, our VIIRS data contain only measurements of the winter
half-year, which are more prone to large changes in albedo. As 
revealed by our analysis, albedo has a strong impact on the NSB.
Thus, data obtained during the winter-half-year tend to show larger scatter.
Moreover, the variation of the satellite's viewing angle adds another factor that increases the scatter.
The monthly averaging may reduce some of these problems (e.g. the viewing angle),
but still large uncertainties as seen in Figure \ref{fig:viirs_trends} remain,
making a long-term trend analysis challenging.
On top of that, the VIIRS/DNB spectral response is very different from the SQM. While the SQM is most sensitive
between 400 and 600nm, the VIIRS/DNB passband covers a range between 500 and 900nm, which is most problematic
in case of a change of the emitted spectral energy distribution towards blue colors (e.g. upgrade to 4000K LEDs).
This should be kept in mind when comparing our SQM-based results to those from VIIRS.
Therefore, we do not expect to see a perfect match between trends obtained from such different instruments.
However, we do expect that direction and magnitude of the trends are comparable.
After conversion to magnitudes, the VIIRS-based trend slopes are 29$\pm$30, 133$\pm$143 and 94$\pm$21~mmag arcsec$^{-2}$ yr$^{-1}$
for BA1, BOD and MAT respectively, i.e. the uncertainties are quite large in relation to our SQM-based study.
Only for one of the stations, MAT, a comparison can be made. Interestingly, in this case, the results do match very well.
From VIIRS an increase of 94$\pm$21~mmag~arcsec$^{-2}$~yr$^{-1}$ (over the DNB bandpass) is measured, while our SQM analysis suggests
a slope of 90$\pm$16 m\magsqm\ yr$^{-1}$. This may be an indication that the color of the night sky at this site has not
changed much.

Finally, we briefly discuss the SQM aging or time-dependent loss of sensitivity as deduced from the twilight method.
Figure \ref{fig:sqm_aging} shows that the correction slopes range from zero to almost $-$80~m\magsqm~yr$^{-1}$ with a mean and
standard deviation of $-$31$\pm$20~m\magsqm\ yr$^{-1}$.
It was previously speculated by \cite{Bara2021} and \cite{Puschnig2021} that the darkening may
be related to the total (time-dependent) amount of (UV) light exposure at a given site.
The scatter of the aging slopes among our stations, however, does not support
such a simple picture. Some other factors, probably dust and polls, may play a crucial role.
Regardless of what causes the loss of sensitivity, we are confident that the post-calibration procedure
via twilight measurements is capable of correcting the change in sensitivity.
An example that provides evidence for the robustness of our method is the comparison of the results
obtained at the stations LGO and LSM. These two SQMs are situated in the same city (Linz), and are thus
only $\sim$1km apart. While for one of the sites (LGO) the aging effect is even below the recognition limit,
for the other one (LSM), an aging slope of $-$35~m\magsqm~yr$^{-1}$ is found. While the correction slopes are quite different,
the final trends are in good agreement. A comparison of several SQMs operating in parallel at different sites within
a small area could potentially provide more insight into the aging effect in the future.

\section{Summary and Conclusion}\label{sec:summary}
We analysed long-term (4-10 years) SQM measurements obtained at 26 different sites,
of which 24 are located in Austria, one in Stockholm (Sweden) and one in Potsdam-Babelsberg (Germany).
We utilize the ``twilight method'' to correct for changing detector sensitivity with
time, i.e. the ``aging effect''. Using aging-corrected data obtained under clear and moonless sky,
long-term trends of light pollution are derived for our sites.
Categorizing our sites into rural, urban and intermediate ones, we find
an average increase in light pollution of approximately 1.7, 1,8 and 3.7 percent per year respectively,
with an estimated $\pm$2-sigma uncertainty of \trenderror\% per year.
The corresponding doubling times are 41, 39 and 19 years.

Furthermore, we establish an empirical atmospheric model which allows us to
investigate the relative impact of 15 different atmospheric parameters
on the night sky brightness.
We find that surface albedo and vegetation have by far the largest impact
on the zenithal night sky brightness. Additionally, black carbon and organic matter aerosols are
important at urban and rural sites, respectively. Snow depth was found to be important for some sites,
while the total column of ozone leaves impact on some rural places.

In the paper we have shown that large-scale (30--40\,km resolution) atmospheric parameters obtained through the
Copernicus Climate Change Service and the Copernicus Atmosphere Monitoring Service
may serve as a basis of a predictive model to constrain the (zenithal) NSB. However, we also
recognized that our empirical model correlates better with SQM observations obtained at rural sites,
i.e. where albedo and vegetation have strongest impact, while aerosols and particulate matter
play only a minor role. Obviously, at urban sites the situation is different. There, aerosols and particulate
matter are typically more abundant, and vary on shorter times and scales.
The weaker correlation between our model and the observations at urban sites
is likely due to the relatively low spatial resolution of the available parameters (aerosol optical depths
and particulate matter). Matching of high-quality aerosol optical depths and
particulate matter would most likely lead to improved predictions also at our urban sites.
This task remains to be done in the future.

\section*{Acknowledgements}
We thank the referee for carefully reading our manuscript and for constructive comments and suggestion.
We acknowledge support from the Alva and Lennart Dahlmark
research grants of the Department of Astronomy at Stockholm University, through which one of the SQMs (STO) and its housing were
financed. We are further thankful to Sergio Gelato (Univ. of Stockholm), who helped with the setup of the SQM in Stockholm (hard-
and software).

Part of this work has been generated using Copernicus Climate Change Service information (2022)
This research made use of SciPy \citep{jones2001scipy} and NumPy \citep{van2011numpy}.

\section*{Data Availability}
The SQM data underlying this article are available for download through repositories
at \url{https://www.land-oberoesterreich.gv.at/115999.htm} and
\url{https://astro.univie.ac.at/en/science-communication/reading-material/light-pollution/}.
Our reduced SQM data can be made available upon reasonable request.

The third party meteorological data used in this article are available through
the Climate Data Store Application Program Interface (CDS API) using
\texttt{Python} (e.g. pip install cdsapi).



\FloatBarrier

\bibliographystyle{mnras}
\bibliography{lp} 

\begin{thebibliography}{}
\makeatletter
\relax
\def\mn@urlcharsother{\let\do\@makeother \do\$\do\&\do\#\do\^\do\_\do\%\do\~}
\def\mn@doi{\begingroup\mn@urlcharsother \@ifnextchar [ {\mn@doi@}
  {\mn@doi@[]}}
\def\mn@doi@[#1]#2{\def\@tempa{#1}\ifx\@tempa\@empty \href
  {http://dx.doi.org/#2} {doi:#2}\else \href {http://dx.doi.org/#2} {#1}\fi
  \endgroup}
\def\mn@eprint#1#2{\mn@eprint@#1:#2::\@nil}
\def\mn@eprint@arXiv#1{\href {http://arxiv.org/abs/#1} {{\tt arXiv:#1}}}
\def\mn@eprint@dblp#1{\href {http://dblp.uni-trier.de/rec/bibtex/#1.xml}
  {dblp:#1}}
\def\mn@eprint@#1:#2:#3:#4\@nil{\def\@tempa {#1}\def\@tempb {#2}\def\@tempc
  {#3}\ifx \@tempc \@empty \let \@tempc \@tempb \let \@tempb \@tempa \fi \ifx
  \@tempb \@empty \def\@tempb {arXiv}\fi \@ifundefined
  {mn@eprint@\@tempb}{\@tempb:\@tempc}{\expandafter \expandafter \csname
  mn@eprint@\@tempb\endcsname \expandafter{\@tempc}}}

\bibitem[\protect\citeauthoryear{Andrei\'{c}}{Andrei\'{c}}{2018}]{Zeljko2018}
Andrei\'{c} v.,  2018, Night sky brightness above Zagreb 2012.-2017,
  \mn@doi{10.48550/ARXIV.1803.00299}, \url {https://arxiv.org/abs/1803.00299}

\bibitem[\protect\citeauthoryear{Bar\'{a}, Lima  \& Zamorano}{Bar\'{a}
  et~al.}{2019}]{Bara2019b}
Bar\'{a} S.,  Lima R.~C.,   Zamorano J.,  2019, \mn@doi [Sustainability]
  {10.3390/su11113070}, 11

\bibitem[\protect\citeauthoryear{Bara, Marco, R\'{i}bas, Garcia, Sanchez~de
  Miguel  \& Zamorano}{Bara et~al.}{2021}]{Bara2021}
Bara S.,  Marco E.,  R\'{i}bas S.,  Garcia M.,  Sanchez~de Miguel A.,
  Zamorano J.,  2021, International Journal of Sustainable Lighting, 23, 1

\bibitem[\protect\citeauthoryear{Bertolo, Binotto, Ortolani  \&
  Sapienza}{Bertolo et~al.}{2019}]{Bertolo2019}
Bertolo A.,  Binotto R.,  Ortolani S.,   Sapienza S.,  2019, \mn@doi [Journal
  of Imaging] {10.3390/jimaging5050056}, 5

\bibitem[\protect\citeauthoryear{{Bessell}}{{Bessell}}{1990}]{Bessel1990}
{Bessell} M.~S.,  1990, \mn@doi [Publications of the Astronomical Society of
  the Pacific] {10.1086/132749}, 102, 1181

\bibitem[\protect\citeauthoryear{{Bessell}}{{Bessell}}{2005}]{Bessel2005}
{Bessell} M.~S.,  2005, \mn@doi [Annual Review of Astronomy and Astrophysics]
  {10.1146/annurev.astro.41.082801.100251}, 43, 293

\bibitem[\protect\citeauthoryear{Bond, Streets, Yarber, Nelson, Woo  \&
  Klimont}{Bond et~al.}{2004}]{Bond2004}
Bond T.~C.,  Streets D.~G.,  Yarber K.~F.,  Nelson S.~M.,  Woo J.-H.,   Klimont
  Z.,  2004, Journal of Geophysical Research: Atmospheres, 109

\bibitem[\protect\citeauthoryear{Boussetta, Balsamo, Beljaars, Kral  \&
  Jarlan}{Boussetta et~al.}{2011}]{Boussetta2011}
Boussetta S.,  Balsamo G.,  Beljaars A.,  Kral T.,   Jarlan L.,  2011, Impact
  of a satellite-derived Leaf Area Index monthly climatology in a global
  Numerical Weather Prediction model, \mn@doi{10.21957/h7n0ilfkp}, \url
  {https://www.ecmwf.int/node/8339}

\bibitem[\protect\citeauthoryear{Boussetta, Balsamo, Dutra, Beljaars  \&
  Albergel}{Boussetta et~al.}{2014}]{Boussetta2014}
Boussetta S.,  Balsamo G.,  Dutra E.,  Beljaars A.,   Albergel C.,  2014,
  Analysis of surface albedo and Leaf Area Index from satellite observations
  and their impact on numerical weather prediction,
  \mn@doi{10.21957/otwcakuu3}, \url {https://www.ecmwf.int/node/12032}

\bibitem[\protect\citeauthoryear{Briegleb \& Ramanathan}{Briegleb \&
  Ramanathan}{1982}]{Briegleb1982}
Briegleb B.,  Ramanathan V.,  1982, \mn@doi [Journal of Applied Meteorology]
  {10.1175/1520-0450(1982)021<1160:SADVIC>2.0.CO;2}, 21, 1160

\bibitem[\protect\citeauthoryear{Briegleb, Minnis, Ramanathan  \&
  Harrison}{Briegleb et~al.}{1986}]{Briegleb1986}
Briegleb B.~P.,  Minnis P.,  Ramanathan V.,   Harrison E.,  1986, \mn@doi
  [Journal of Climate and Applied Meteorology]
  {10.1175/1520-0450(1986)025<0214:CORCSA>2.0.CO;2}, 25, 214

\bibitem[\protect\citeauthoryear{Cao, Liang, Tian, Zhang, Quan  \& Liu}{Cao
  et~al.}{2014}]{Cao2014}
Cao X.,  Liang J.,  Tian P.,  Zhang L.,  Quan X.,   Liu W.,  2014, \mn@doi
  [Atmospheric Pollution Research] {https://doi.org/10.5094/APR.2014.069}, 5,
  601

\bibitem[\protect\citeauthoryear{{Cavazzani}, {Ortolani}, {Bertolo}, {Binotto},
  {Fiorentin}, {Carraro}, {Saviane}  \& {Zitelli}}{{Cavazzani}
  et~al.}{2020}]{Cavazzani2020}
{Cavazzani} S.,  {Ortolani} S.,  {Bertolo} A.,  {Binotto} R.,  {Fiorentin} P.,
  {Carraro} G.,  {Saviane} I.,   {Zitelli} V.,  2020, \mn@doi [Monthly Notices
  of the Royal Astronomical Society] {10.1093/mnras/staa416}, \href
  {https://ui.adsabs.harvard.edu/abs/2020MNRAS.493.2463C} {493, 2463}

\bibitem[\protect\citeauthoryear{Chepesiuk}{Chepesiuk}{2009}]{Chepesiuk2009}
Chepesiuk R.,  2009, \mn@doi [Environmental health perspectives]
  {10.1289/ehp.117-a20}, 117, A20

\bibitem[\protect\citeauthoryear{Cho, Ryu, Lee, Kim, Lee  \& Choi}{Cho
  et~al.}{2015}]{Cho2015}
Cho Y.,  Ryu S.-H.,  Lee B.,  Kim K.,  Lee E.,   Choi J.,  2015, \mn@doi
  [Chronobiology international] {10.3109/07420528.2015.1073158}, 32, 1

\bibitem[\protect\citeauthoryear{{Cinzano}}{{Cinzano}}{2005}]{Cinzano2005}
{Cinzano} P.,  2005, Technical report, {Night Sky Photometry with Sky Quality
  Meter}, \url {http://www.inquinamentoluminoso.it/download/sqmreport.pdf}.
Dipartimento di Astronomia, Vicolo dell Osservatorio 2, I-35100 Padova, Italy,
  Istituto di Scienza e Tecnologia dell Inquinamento Luminoso, Via Roma 13,
  I-36106 Thiene, Italy, \url
  {http://www.inquinamentoluminoso.it/download/sqmreport.pdf}

\bibitem[\protect\citeauthoryear{Coakley}{Coakley}{2003}]{Coakley2003}
Coakley J.,  2003, in Holton J.~R.,  ed., , Encyclopedia of Atmospheric
  Sciences.
Academic Press, Oxford, pp 1914 -- 1923,
  \mn@doi{https://doi.org/10.1016/B0-12-227090-8/00069-5}, \url
  {http://www.sciencedirect.com/science/article/pii/B0122270908000695}

\bibitem[\protect\citeauthoryear{Eisenbeis}{Eisenbeis}{2006}]{Eisenbeis2006}
Eisenbeis G.,  2006, Ecological consequences of artificial night lighting.
Island Press, pp 281--304

\bibitem[\protect\citeauthoryear{Garcia-Saenz et~al.,}{Garcia-Saenz
  et~al.}{2018}]{Garcia-Saenz2018}
Garcia-Saenz A.,  et~al., 2018, \mn@doi [Environmental Health Perspectives]
  {10.1289/EHP1837}, 126, 047011

\bibitem[\protect\citeauthoryear{Haim \& Portnov}{Haim \&
  Portnov}{2013}]{Haim2013}
Haim A.,  Portnov B.~A.,  2013, Light Pollution as a New Risk Factor for Human
  Breast and Prostate Cancers, 1st ed. 2013. edn.
Springer Netherlands

\bibitem[\protect\citeauthoryear{H{\"a}nel et~al.,}{H{\"a}nel
  et~al.}{2018}]{Haenel2018}
H{\"a}nel A.,  et~al., 2018, \mn@doi [Journal of Quantitative Spectroscopy and
  Radiative Transfer] {https://doi.org/10.1016/j.jqsrt.2017.09.008}, 205, 278

\bibitem[\protect\citeauthoryear{Hersbach et~al.,}{Hersbach
  et~al.}{2018}]{ERA5}
Hersbach H.,  et~al., 2018, Operational global reanalysis: progress, future
  directions and synergies with NWP, \mn@doi{10.21957/tkic6g3wm}, \url
  {https://www.ecmwf.int/node/18765}

\bibitem[\protect\citeauthoryear{H\"olker, Wolter  \& Perkin}{H\"olker
  et~al.}{2010}]{Hoelker2010}
H\"olker F.,  Wolter C.,   Perkin E.,  2010, \mn@doi [Trends in ecology &
  evolution] {10.1016/j.tree.2010.09.007}, 25, 681

\bibitem[\protect\citeauthoryear{Inness et~al.,}{Inness et~al.}{2019}]{CAMS}
Inness A.,  et~al., 2019, \mn@doi [Atmos. Chem. Phys.]
  {10.5194/acp-19-3515-2019}, 19, 3515

\bibitem[\protect\citeauthoryear{Jechow, H{\"o}lker  \& Kyba}{Jechow
  et~al.}{2019}]{Jechow2019}
Jechow A.,  H{\"o}lker F.,   Kyba C. C.~M.,  2019, \mn@doi [Scientific Reports]
  {10.1038/s41598-018-37817-8}, 9, 1391

\bibitem[\protect\citeauthoryear{{Jones}, Oliphant, Peterson  \&
  Others}{{Jones} et~al.}{2001}]{jones2001scipy}
{Jones} E.,  Oliphant T.,  Peterson P.,   Others 2001, {SciPy}: Open source
  scientific tools for Python, \url {http://www.scipy.org/}

\bibitem[\protect\citeauthoryear{{Jones}, {Noll}, {Kausch}, {Szyszka}  \&
  {Kimeswenger}}{{Jones} et~al.}{2013}]{Jones2013}
{Jones} A.,  {Noll} S.,  {Kausch} W.,  {Szyszka} C.,   {Kimeswenger} S.,  2013,
  \mn@doi [\aap] {10.1051/0004-6361/201322433}, 560, A91

\bibitem[\protect\citeauthoryear{Khodasevich, Tsui, Keung, Skene  \&
  Martinez}{Khodasevich et~al.}{2020}]{Khodasevich2020}
Khodasevich D.,  Tsui S.,  Keung D.,  Skene D.~J.,   Martinez M.~E.,  2020,
  \mn@doi [medRxiv] {10.1101/2020.10.21.20214676}

\bibitem[\protect\citeauthoryear{Kocifaj \& Barentine}{Kocifaj \&
  Barentine}{2021}]{Kocifaj2021}
Kocifaj M.,  Barentine J.~C.,  2021, \mn@doi [Scientific Reports]
  {10.1038/s41598-021-94241-1}, 11, 14622

\bibitem[\protect\citeauthoryear{Kyba, Ruhtz, Fischer  \& H\"{o}lker}{Kyba
  et~al.}{2011}]{Kyba2011}
Kyba C. C.~M.,  Ruhtz T.,  Fischer J.,   H\"{o}lker F.,  2011, \mn@doi [PLOS
  ONE] {10.1371/journal.pone.0017307}, 6, 1

\bibitem[\protect\citeauthoryear{Kyba et~al.,}{Kyba et~al.}{2015}]{Kyba2015}
Kyba C. C.~M.,  et~al., 2015, \mn@doi [Scientific Reports] {10.1038/srep08409},
  5

\bibitem[\protect\citeauthoryear{Longcore \& Rich}{Longcore \&
  Rich}{2004}]{Longcore2004}
Longcore T.,  Rich C.,  2004, \mn@doi [Frontiers in Ecology and the
  Environment] {10.1890/1540-9295(2004)002[0191:ELP]2.0.CO;2}, 2, 191

\bibitem[\protect\citeauthoryear{Masana, Carrasco, Bará  \& Ribas}{Masana
  et~al.}{2020}]{Masana2020}
Masana E.,  Carrasco J.~M.,  Bará S.,   Ribas S.~J.,  2020, \mn@doi [Monthly
  Notices of the Royal Astronomical Society] {10.1093/mnras/staa4005}, 501,
  5443

\bibitem[\protect\citeauthoryear{Masana, Bará, Carrasco  \& Ribas}{Masana
  et~al.}{2022}]{Masana2022}
Masana E.,  Bará S.,  Carrasco J.~M.,   Ribas S.~J.,  2022, International
  Journal of Sustainable Lighting, 24

\bibitem[\protect\citeauthoryear{Mathews, Roche, Aughney, Jones, Day, Baker  \&
  Langton}{Mathews et~al.}{2015}]{Mathews2015}
Mathews F.,  Roche N.,  Aughney T.,  Jones N.,  Day J.,  Baker J.,   Langton
  S.,  2015, \mn@doi [Philosophical Transactions of the Royal Society B:
  Biological Sciences] {10.1098/rstb.2014.0124}, 370, 20140124

\bibitem[\protect\citeauthoryear{Men{\'e}ndez-Vel{\'a}zquez, Morales  \&
  Garc{\'\i}a-Delgado}{Men{\'e}ndez-Vel{\'a}zquez
  et~al.}{2022}]{Menendez-Velazquez2022}
Men{\'e}ndez-Vel{\'a}zquez A.,  Morales D.,   Garc{\'\i}a-Delgado A.~B.,  2022,
  Int J Environ Res Public Health, 19

\bibitem[\protect\citeauthoryear{{Noll}, {Kausch}, {Barden}, {Jones},
  {Szyszka}, {Kimeswenger}  \& {Vinther}}{{Noll} et~al.}{2012}]{Noll2012}
{Noll} S.,  {Kausch} W.,  {Barden} M.,  {Jones} A.~M.,  {Szyszka} C.,
  {Kimeswenger} S.,   {Vinther} J.,  2012, \mn@doi [\aap]
  {10.1051/0004-6361/201219040}, 543, A92

\bibitem[\protect\citeauthoryear{Owens, Cochard, Durrant, Farnworth, Perkin  \&
  Seymoure}{Owens et~al.}{2020}]{Owens2020}
Owens A.~C.,  Cochard P.,  Durrant J.,  Farnworth B.,  Perkin E.~K.,   Seymoure
  B.,  2020, \mn@doi [Biological Conservation]
  {https://doi.org/10.1016/j.biocon.2019.108259}, 241, 108259

\bibitem[\protect\citeauthoryear{Parkinson, Lawson  \& Tiegs}{Parkinson
  et~al.}{2020}]{Parkinson2020}
Parkinson E.,  Lawson J.,   Tiegs S.~D.,  2020, \mn@doi [PLOS ONE]
  {10.1371/journal.pone.0240138}, 15, 1

\bibitem[\protect\citeauthoryear{{Patat}, {Ugolnikov}  \&
  {Postylyakov}}{{Patat} et~al.}{2006}]{Patat2006}
{Patat} F.,  {Ugolnikov} O.~S.,   {Postylyakov} O.~V.,  2006, \mn@doi
  [Astronomy and Astrophysics] {10.1051/0004-6361:20064992}, \href
  {https://ui.adsabs.harvard.edu/abs/2006A&A...455..385P} {455, 385}

\bibitem[\protect\citeauthoryear{Perkin, H\"{o}lker, Richardson, Sadler, Wolter
   \& Tockner}{Perkin et~al.}{2011}]{Perkin2011}
Perkin E.~K.,  H\"{o}lker F.,  Richardson J.~S.,  Sadler J.~P.,  Wolter C.,
  Tockner K.,  2011, \mn@doi [Ecosphere]
  {https://doi.org/10.1890/ES11-00241.1}, 2, art122

\bibitem[\protect\citeauthoryear{Perkin, H\"{o}lker  \& Tockner}{Perkin
  et~al.}{2014}]{Perkin2013}
Perkin E.~K.,  H\"{o}lker F.,   Tockner K.,  2014, \mn@doi [Freshwater Biology]
  {https://doi.org/10.1111/fwb.12270}, 59, 368

\bibitem[\protect\citeauthoryear{{Posch}, {Binder}  \& {Puschnig}}{{Posch}
  et~al.}{2018}]{Posch2018}
{Posch} T.,  {Binder} F.,   {Puschnig} J.,  2018, \mn@doi [Journal of
  Quantitative Spectroscopy and Radiative Transfer]
  {10.1016/j.jqsrt.2018.03.010}, \href
  {https://ui.adsabs.harvard.edu/abs/2018JQSRT.211..144P} {211, 144}

\bibitem[\protect\citeauthoryear{{Puschnig}, {Posch}  \&
  {Uttenthaler}}{{Puschnig} et~al.}{2014a}]{Puschnig2014a}
{Puschnig} J.,  {Posch} T.,   {Uttenthaler} S.,  2014a, \mn@doi [Journal of
  Quantitative Spectroscopy and Radiative Transfer]
  {10.1016/j.jqsrt.2013.08.019}, \href
  {https://ui.adsabs.harvard.edu/abs/2014JQSRT.139...64P} {139, 64}

\bibitem[\protect\citeauthoryear{{Puschnig}, {Schwope}, {Posch}  \&
  {Schwarz}}{{Puschnig} et~al.}{2014b}]{Puschnig2014b}
{Puschnig} J.,  {Schwope} A.,  {Posch} T.,   {Schwarz} R.,  2014b, \mn@doi
  [Journal of Quantitative Spectroscopy and Radiative Transfer]
  {10.1016/j.jqsrt.2013.12.011}, \href
  {https://ui.adsabs.harvard.edu/abs/2014JQSRT.139...76P} {139, 76}

\bibitem[\protect\citeauthoryear{{Puschnig}, {Wallner}  \& {Posch}}{{Puschnig}
  et~al.}{2020}]{Puschnig2020}
{Puschnig} J.,  {Wallner} S.,   {Posch} T.,  2020, \mn@doi [\mnras]
  {10.1093/mnras/stz3514}, \href
  {https://ui.adsabs.harvard.edu/abs/2020MNRAS.492.2622P} {492, 2622}

\bibitem[\protect\citeauthoryear{{Puschnig}, {N{\"a}slund}, {Schwope}  \&
  {Wallner}}{{Puschnig} et~al.}{2021}]{Puschnig2021}
{Puschnig} J.,  {N{\"a}slund} M.,  {Schwope} A.,   {Wallner} S.,  2021, \mn@doi
  [Monthly Notices of the Royal Astronomical Society] {10.1093/mnras/staa4019},
  \href {https://ui.adsabs.harvard.edu/abs/2021MNRAS.502.1095P} {502, 1095}

\bibitem[\protect\citeauthoryear{Roesch, Wild, Pinker  \& Ohmura}{Roesch
  et~al.}{2002}]{Roesch2002}
Roesch A.,  Wild M.,  Pinker R.,   Ohmura A.,  2002, \mn@doi [Journal of
  Geophysical Research: Atmospheres] {10.1029/2001JD000809}, 107, ACL 13

\bibitem[\protect\citeauthoryear{Schmidt \& Spoelstra}{Schmidt \&
  Spoelstra}{2020}]{Schmidt2020}
Schmidt W.,  Spoelstra H.,  2020, Darkness monitoring in the Netherlands
  2009-2019, \mn@doi{10.5281/zenodo.4293366}, \url
  {https://doi.org/10.5281/zenodo.4293366}

\bibitem[\protect\citeauthoryear{Schwarz}{Schwarz}{1978}]{Schwarz1987}
Schwarz G.,  1978, \mn@doi [The Annals of Statistics] {10.1214/aos/1176344136},
  6, 461

\bibitem[\protect\citeauthoryear{Schwarz et~al.,}{Schwarz
  et~al.}{2006}]{Schwarz2006}
Schwarz J.~P.,  et~al., 2006, Journal of Geophysical Research: Atmospheres, 111

\bibitem[\protect\citeauthoryear{\'{S}ci\k{e}\.{z}or}{\'{S}ci\k{e}\.{z}or}{2020}]{Sciezor2020a}
\'{S}ci\k{e}\.{z}or T.,  2020, \mn@doi [Journal of Quantitative Spectroscopy
  and Radiative Transfer] {https://doi.org/10.1016/j.jqsrt.2020.106962}, 247,
  106962

\bibitem[\protect\citeauthoryear{\'{S}ci\k{e}\.{z}or \&
  Czaplicka}{\'{S}ci\k{e}\.{z}or \& Czaplicka}{2020}]{Sciezor2020b}
\'{S}ci\k{e}\.{z}or T.,  Czaplicka A.,  2020, \mn@doi [Journal of Quantitative
  Spectroscopy and Radiative Transfer]
  {https://doi.org/10.1016/j.jqsrt.2020.107168}, 254, 107168

\bibitem[\protect\citeauthoryear{\'{S}ci\k{e}\.{z}or \&
  Kubala}{\'{S}ci\k{e}\.{z}or \& Kubala}{2014}]{Sciezor2014}
\'{S}ci\k{e}\.{z}or T.,  Kubala M.,  2014, \mn@doi [Monthly Notices of the
  Royal Astronomical Society] {10.1093/mnras/stu1577}, 444, 2487

\bibitem[\protect\citeauthoryear{Tibshirani}{Tibshirani}{1996}]{lasso}
Tibshirani R.,  1996, \mn@doi [Journal of the Royal Statistical Society: Series
  B (Methodological)] {https://doi.org/10.1111/j.2517-6161.1996.tb02080.x}, 58,
  267

\bibitem[\protect\citeauthoryear{Van Der~Walt, Colbert  \& Varoquaux}{Van
  Der~Walt et~al.}{2011}]{van2011numpy}
Van Der~Walt S.,  Colbert S.~C.,   Varoquaux G.,  2011, Computing in Science \&
  Engineering, 13, 22

\bibitem[\protect\citeauthoryear{{Wallner} \& {Kocifaj}}{{Wallner} \&
  {Kocifaj}}{2019}]{Wallner2019}
{Wallner} S.,  {Kocifaj} M.,  2019, \mn@doi [\jqsrt]
  {10.1016/j.jqsrt.2019.106648}, \href
  {https://ui.adsabs.harvard.edu/abs/2019JQSRT.23906648W} {239, 106648}

\bibitem[\protect\citeauthoryear{Zamorano et~al.,}{Zamorano
  et~al.}{2015}]{Zamorano2015}
Zamorano J.,  et~al., 2015, IAU General Assembly, 22, 2254642

\makeatother
\end{thebibliography}


\appendix

\section{A synthetic sky model for the SQM band as a Reference}\label{sec:appendix_skymodel}
In order to quantify anthropogenic light at night, knowledge of the natural night sky brightness is needed.
An all-sky model that takes into account scattered moonlight, starlight, molecular emission of the lower atmosphere,
emission lines in the upper atmosphere and airglow, was published by \cite{Noll2012} and \cite{Jones2013},
as part of an Austrian in-kind contribution to the European Southern Observatory (ESO), e.g. ESO's exposure time
calculator is based on the model.

The main input parameters are zenith distance or airmass of the observation, precipitable water vapor and
monthly averaged solar flux. For the moon radiance component, the separation of Sun and Moon as seen from Earth, the
Moon-target separation, Moon altitude over horizon and the Moon-Earth distance are needed.

The result is a synthetic (cloud-free) night sky spectrum for the target location.

For our purpose, we
multiply
the so derived night sky spectrum with the SQM transmission curve
and calculate a synthetic SQM magnitude
via integration.
We have decided to make some simplifications,
allowing us to evaluate the model on a 2-dimensional parameter grid with
vectors of (Sun-Moon-separation, Moon altitude) only.
This is reasonable in our case, because the measurement devices we are using, the Sky Quality Meters of type SQM-LE,
are equipped with a front lens that narrows down the field of view to a roughly 20 degree wide cone, pointed towards zenith.
Hence, we only need to consider zenithal night sky brightness. The two input parameters \textit{Moon-target separation} and \textit{Moon altitude}
can thus be simplified to one parameter, with the former one being the \textit{Moon zenith distance}. We have further decided to evaluate
the model for a fixed precipitable water vapor value of 5\,mm, a monthly averaged solar flux of 130\,sfu and for a fixed mean Moon-Earth distance.
These simplifications have practically no influence on our results, since ALAN's contribution to our SQM measurements is magnitudes
larger than the natural variation caused by phenomena such as water vapor or solar flux. However, variations due to Moon
phase and height are large enough to be important and are thus fully treated by our gridded model evaluation for the zenith.

Since the natural, cloudless sky brightness changes smoothly, a grid spacing of one degree in both parameters (Sun-Moon-separation, Moon altitude)
was found to be sufficient.

\begin{figure}
        \begin{center}
        \includegraphics[width=1\columnwidth]{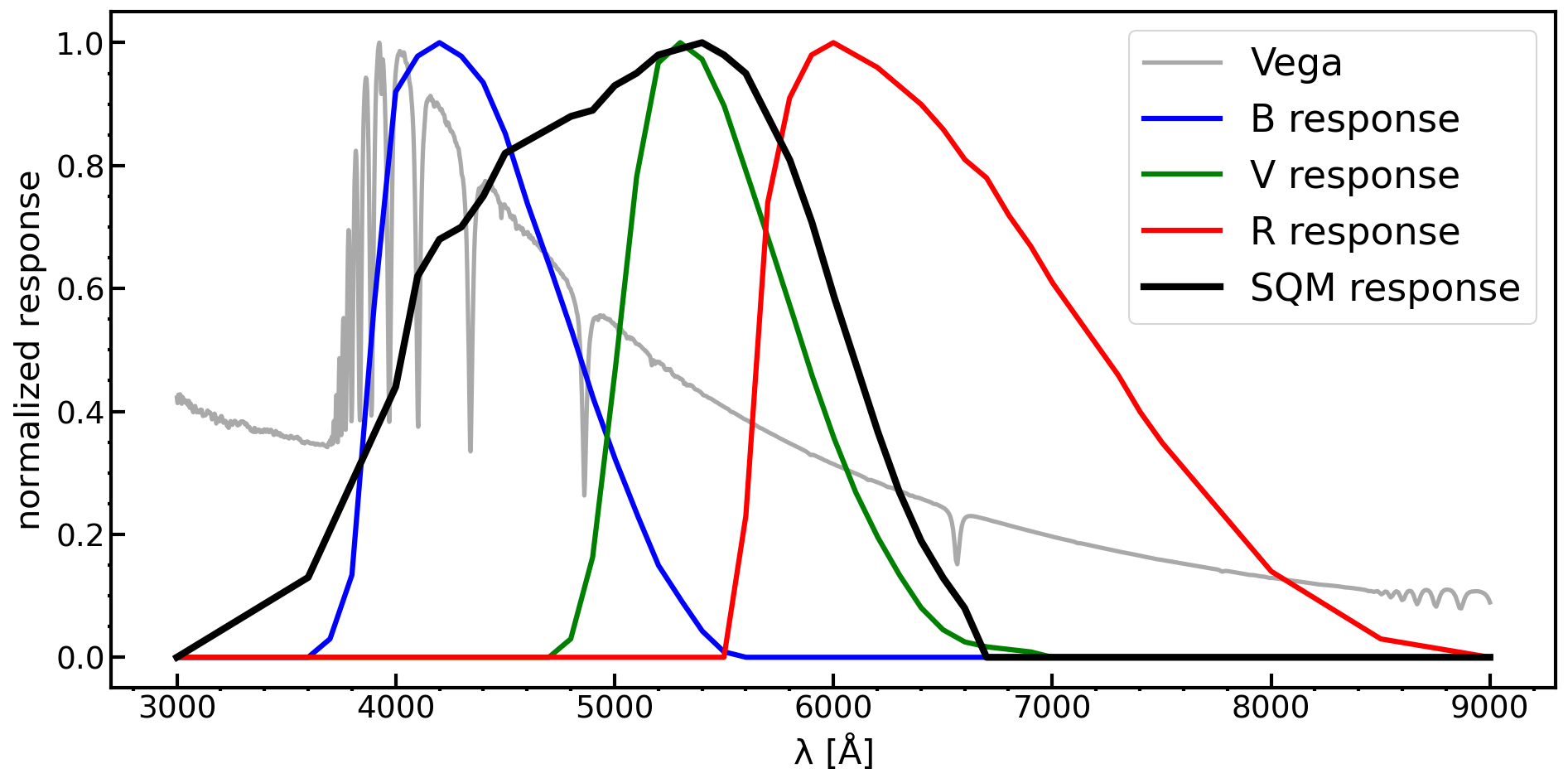}
        \caption[transmission curves]{Bessel B, V, R and SQM transmission curves (\textit{solid curves}) and Gaussian fit results (\textit{dashed curves})
        on top of the spectrum of Vega.}
        \label{fig:sqmtrans}
        \end{center}
\end{figure}

The Sky Quality Meters (SQMs) are produced by Unihedron\footnote{http://www.unihedron.com/}.
Several types are available, of which we are using those denoted by SQM-LE, indicating that they are equipped with a lens (L),
and connected via ethernet (E).
The lens narrows down the field of view to a half-width-half-maximum (HWHM) angular sensitivity of 10 degree, corresponding to
a cone with an opening angle of 20 degree.
The spectral sensitivity of the SQM is sometimes compared to the V-band, but as shown in Figure \ref{fig:sqmtrans}
it is much wider towards the blue end.
Detailed technical specifications are found in \cite{Cinzano2005}.

All our SQMs are operated inside weather-proof housings, also produced by Unihedron. Our measurements are corrected
for the loss of light due to the housing's cover glass (via subtraction of 0.11\,\magsqm\ from the reading).

We aim to compare our SQM measurements to natural (cloud-free) night sky brightness values derived
through a sky model. From our gridded model implementation, spectra in physical flux units of $\mathrm{photons}\ s^{-1}\ m^{-2}\ \mathrm{\upmu m}^{-1}\ \mathrm{arcsec}^{-2}$
are obtained, which we convert to $\mathrm{erg}\ s^{-1}\ \mathrm{cm}^{-2}\ \angstrom^{-1}\ \mathrm{arcsec}^{-2}$.
The spectra are then
multiplied
with the SQM transmission curve. Subsequently, the total flux is calculated via integration and the result is divided by the bandwidth,
which finally gives again a flux density in $\mathrm{erg}\ s^{-1}\ \mathrm{cm}^{-2}\ \angstrom^{-1}\ \mathrm{arcsec}^{-2}$.
The transformation into a Vega-based SQM magnitude (that should be compared to our measurements),
is done using the \textit{photometric zeropoint} and the \textit{zero magnitude flux} for the SQM. The latter two quantities were derived
as explained in the following:

\begin{itemize}
\item A Vega spectrum was downloaded from the \textit{CALSPEC Calibration Database}\footnote{https://archive.stsci.edu/hlsps/reference-atlases/cdbs/current\_calspec/}.
The flux density of the spectrum is given in units of \fluxunit.
For Vega (A0V star) the V-band magnitude is 0.03\,mag and its (B-V) color is 0.

\item The Vega spectrum is
multiplied
with the SQM transmission curve, and for sanity checks also with B, V, R Bessel filter transmission curves using data from \cite{Bessel1990}.

\item The bandwidths were calculated
via integration of the filter curves along the wavelength axis (see Figure \ref{fig:sqmtrans} for the transmission curves used).
As seen in Table \ref{tab:filter}, our results
agree on a 10-percent level with those published by \cite{Masana2022}, who found 959, 909, 1634 and 2228\,\AA\ for B, V, R and the SQM band.

\item The integrated flux of the
product (Vega*Filter)
is divided by the bandwidth. The resulting
flux density (FD) is then given in units of \fluxunit. For B, V, R the zeropoints (ZP) are calculated via:
$0.03=-2.5\mathrm{log(FD)+ZP}$.
\cite{Bessel2005} published zeropoints for these filters. They are -20.45, -21.12 and -21.61\,mag for B, V and R respectively.
Our results agree at a level of 0.1\,mag, which is better than the absolute photometric accuracy of the SQM.

\item It is known that Unihedron's in-house calibration of the SQM is performed using a green LED. This calibration
leads to (SQM$-$V)\,=\,$-$0.35 \citep[see][Figure 15]{Cinzano2005}. Thus, the SQM \textit{photometric zeropoint} is calculated
via: $-0.32=-2.5\mathrm{log(FD)+ZP}$. The SQM zeropoint is -21.18$\pm$0.1\,mag. The zeropoint is defined as the magnitude of a source having unity flux.

\item
Finally, the \textit{zero magnitude flux} (FD$_0$) can be calculated using the zeropoint from above and: $0=-2.5\mathrm{log(FD_0)+ZP}$. For the
SQM the zero magnitude flux is $3.38\ \times\ 10^{-9}$ \fluxunit.
\end{itemize}

\begin{table}
\centering
\caption{Filter (\textit{column 1}), peak wavelength (\textit{column 2}) and bandwidth (\textit{column 3}).
The zeropoint (ZP), defined as the magnitude of a source having a flux of 1\,\fluxunit (\textit{column 4}), and the zero-magnitude flux (\textit{column 5}),
were derived
via multiplication with subsequent integration
of a Vega spectrum with the corresponding filter transmission curve.}
\label{tab:filter}
\begin{tabular}{ l | c c c c}
Filter   &   $\lambda_{peak}$   &    BW     &    ZP       &     $FD_{0}$   \\
         &   [$\angstrom$]            &    [$\angstrom$]  &    [$\mathrm{mag}$]  &     [$10^{-9}$ \fluxunit] \\
\hline \hline 
B        &   4350               &    959      &    -20.48   &     6.43 \\
V        &   5437               &    893      &    -21.08   &     3.69 \\
R        &   6433               &   1591      &    -21.64   &     2.21 \\
SQM      &   5087               &   2008      &    -21.18   &     3.38 \\
\end{tabular}
\end{table}

\section{Long-term trends}

\begin{figure*}
\centering
        \includegraphics[width=1\columnwidth]{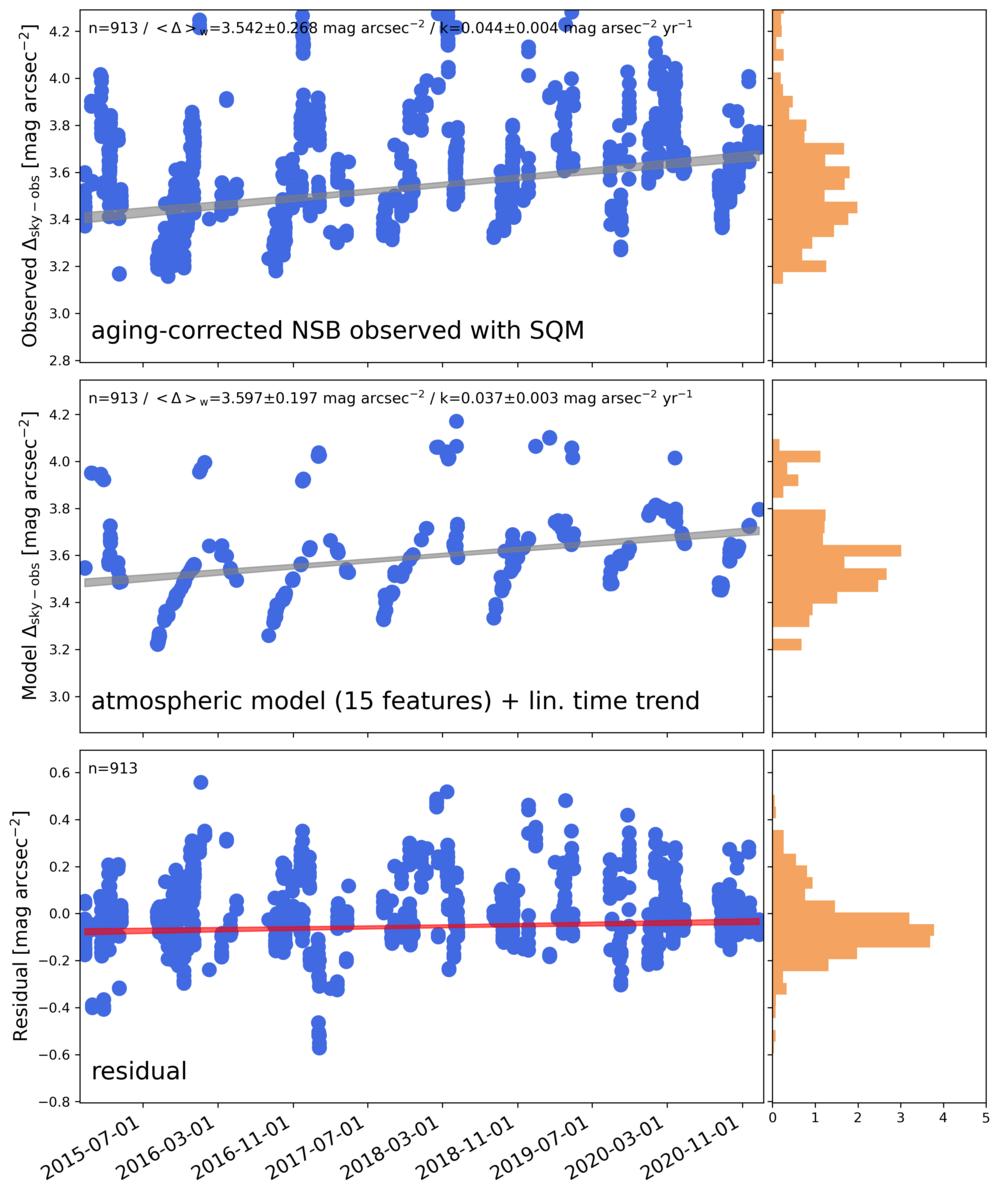}
        \includegraphics[width=1\columnwidth]{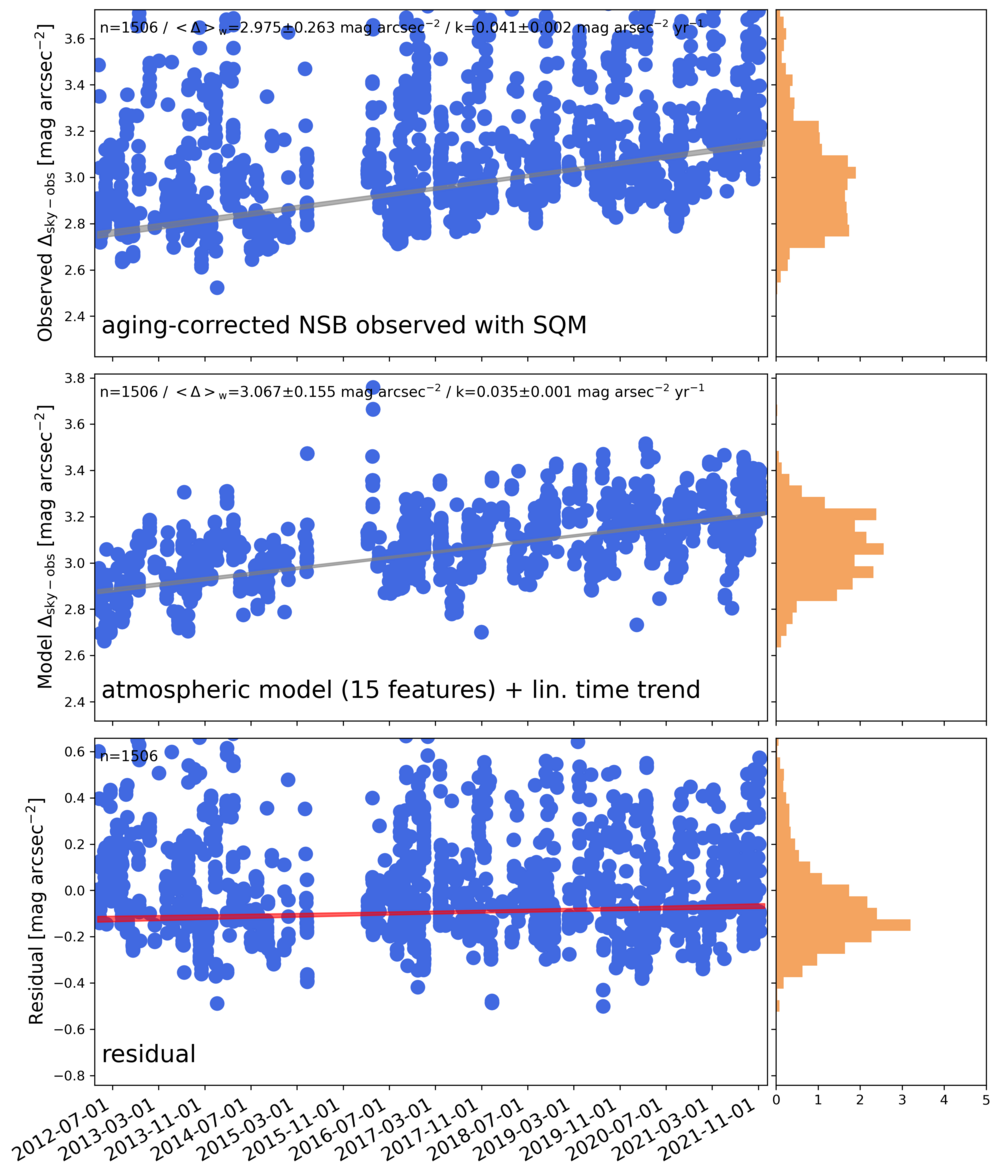}
        \caption[Long-term Trendanalysis]{Long-term trend for STO and IFA. See caption Figure \ref{fig:BA1_BOD_trend} for more details.}
        \label{fig:STO_IFA_trend}
\end{figure*}

\begin{figure*}
\centering
        \includegraphics[width=1\columnwidth]{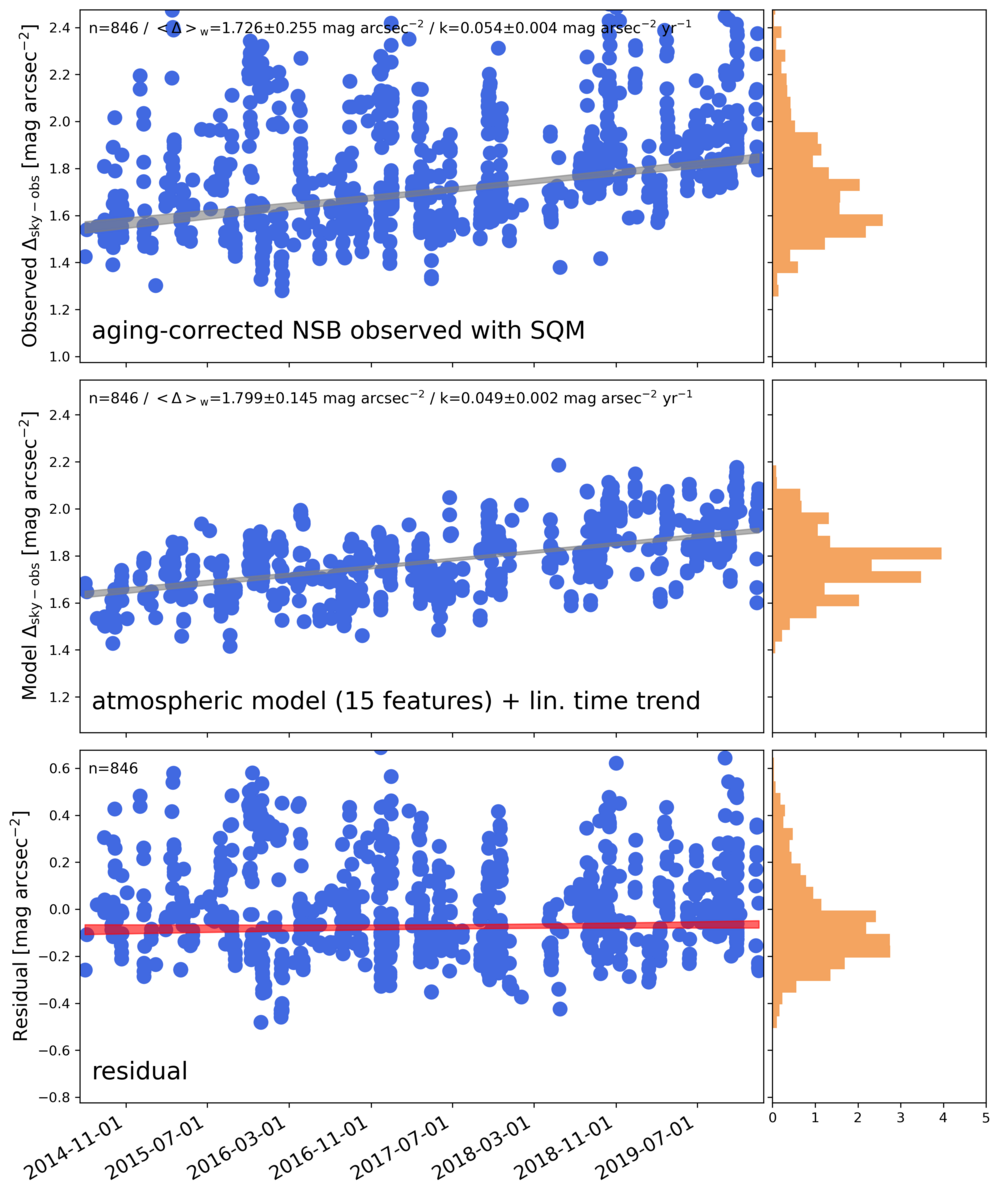}
        \includegraphics[width=1\columnwidth]{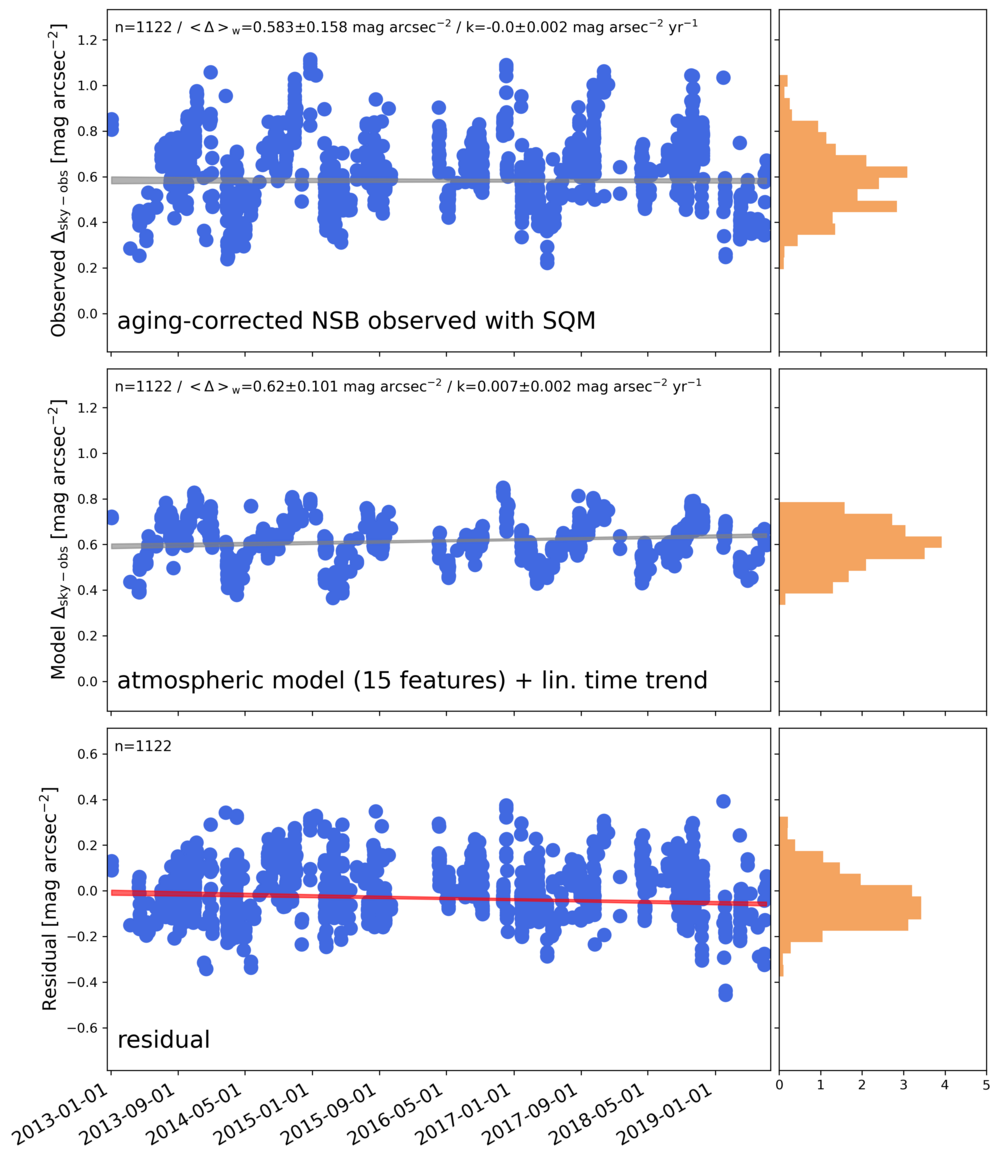}
        \caption[Long-term Trendanalysis]{Long-term trend for GRA and FOA. See caption Figure \ref{fig:BA1_BOD_trend} for more details.}
        \label{fig:GRA_FOA_trend}
\end{figure*}

\begin{figure*}
\centering
        \includegraphics[width=1\columnwidth]{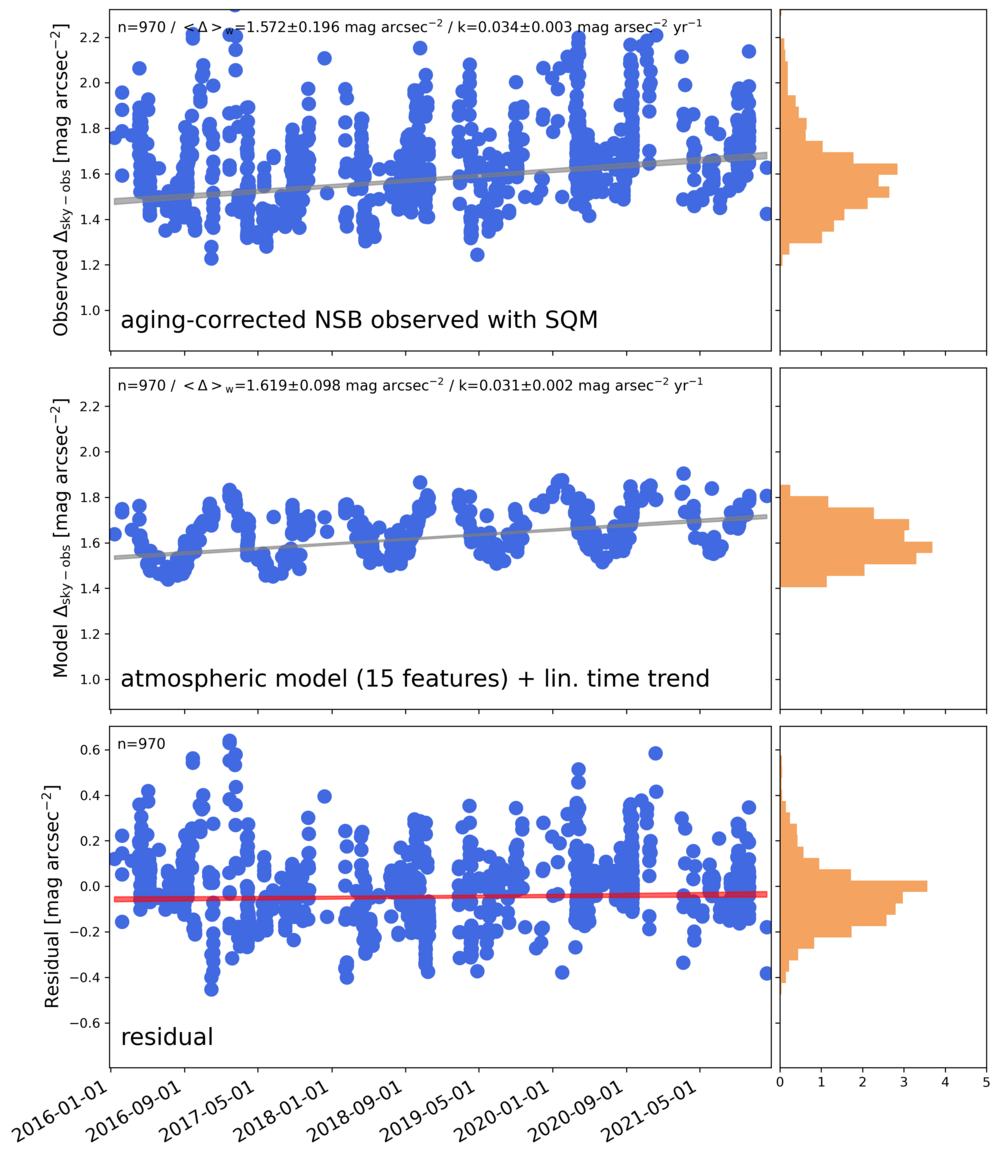}
        \includegraphics[width=1\columnwidth]{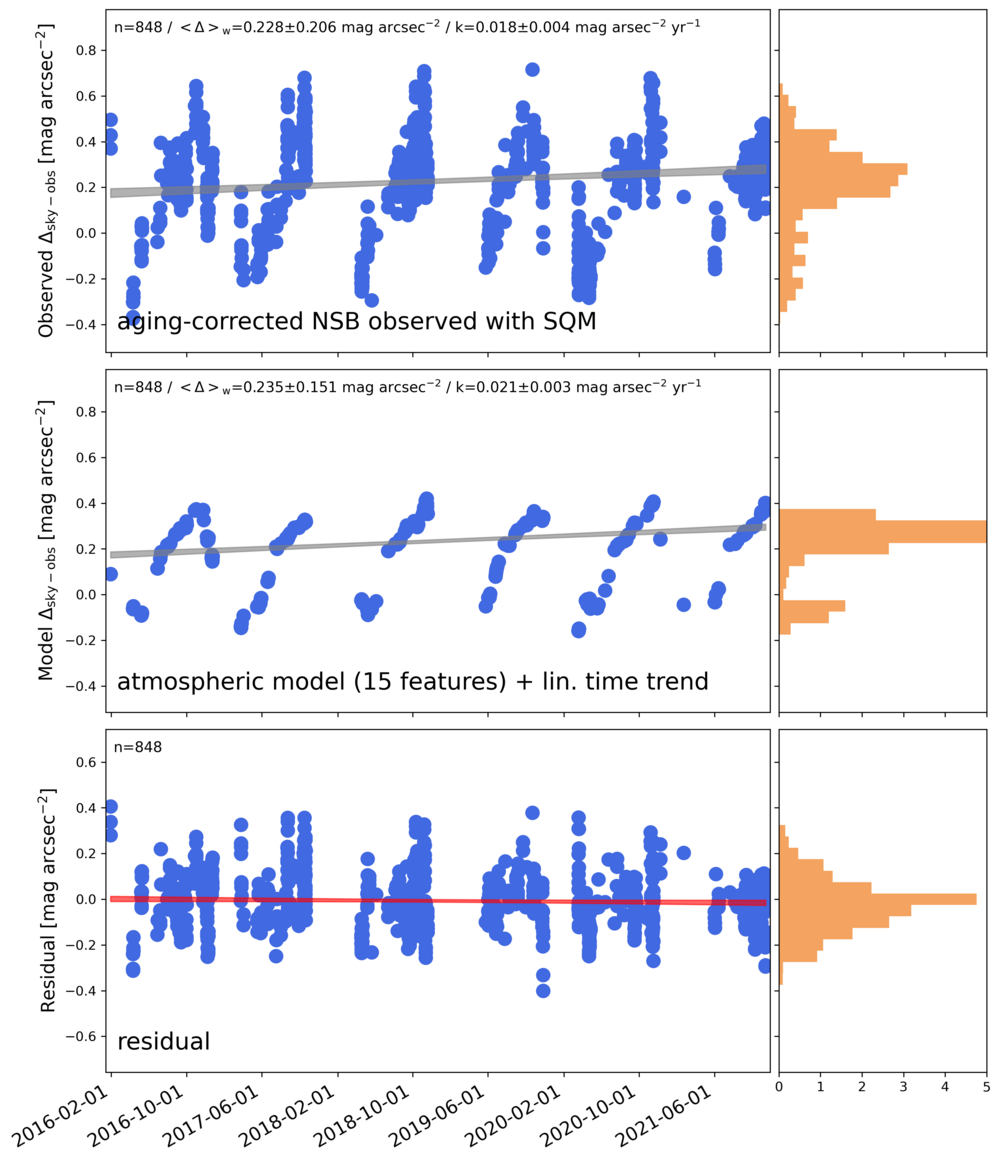}
        \caption[Long-term Trendanalysis]{Long-term trend for BRA and FEU. See caption Figure \ref{fig:BA1_BOD_trend} for more details.}
        \label{fig:BRA_FEU_trend}
\end{figure*}

\begin{figure*}
\centering
        \includegraphics[width=1\columnwidth]{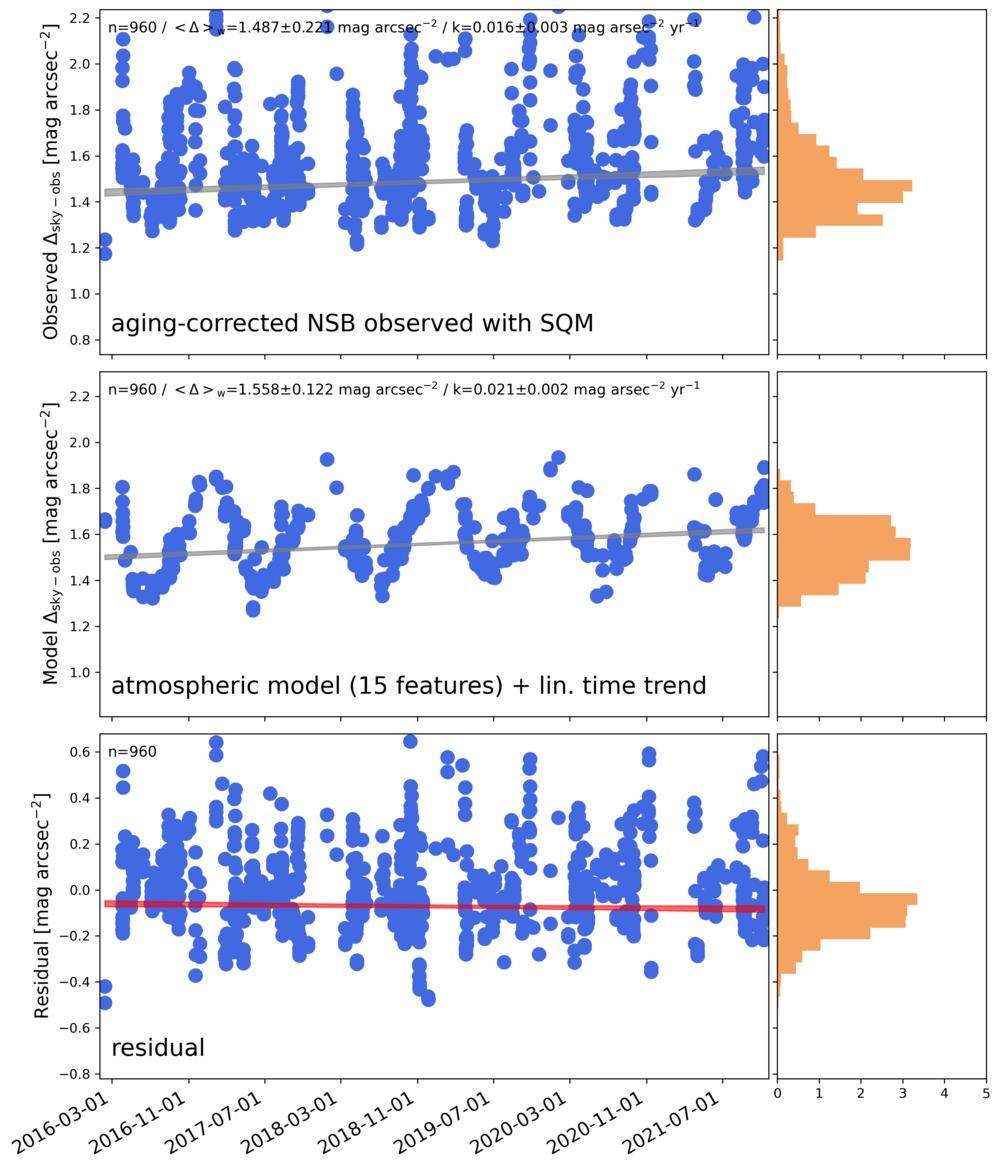}
        \includegraphics[width=1\columnwidth]{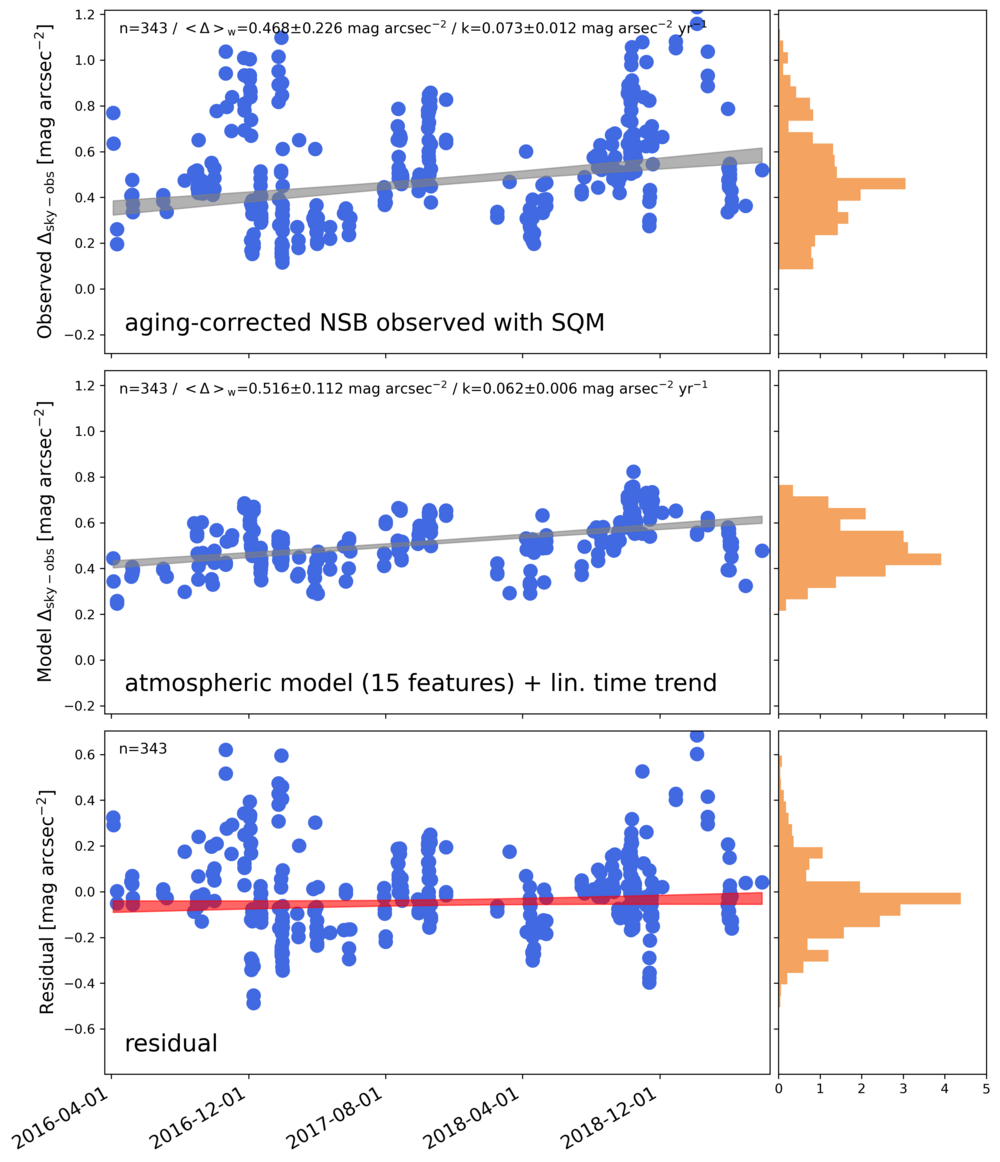}
        \caption[Long-term Trendanalysis]{Long-term trend for FRE and GIS. See caption Figure \ref{fig:BA1_BOD_trend} for more details.}
        \label{fig:FRE_GIS_trend}
\end{figure*}

\begin{figure*}
\centering
        \includegraphics[width=1\columnwidth]{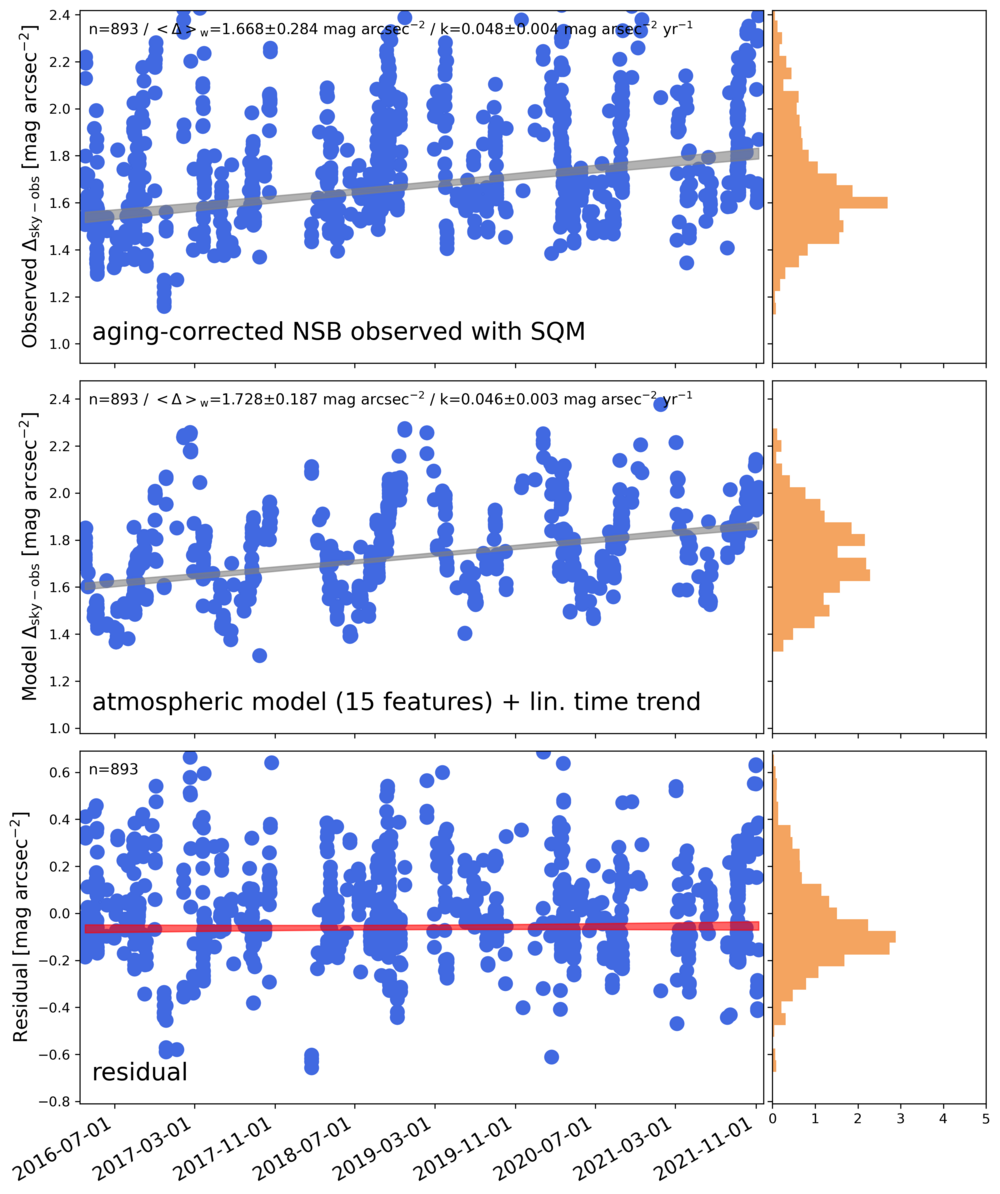}
        \includegraphics[width=1\columnwidth]{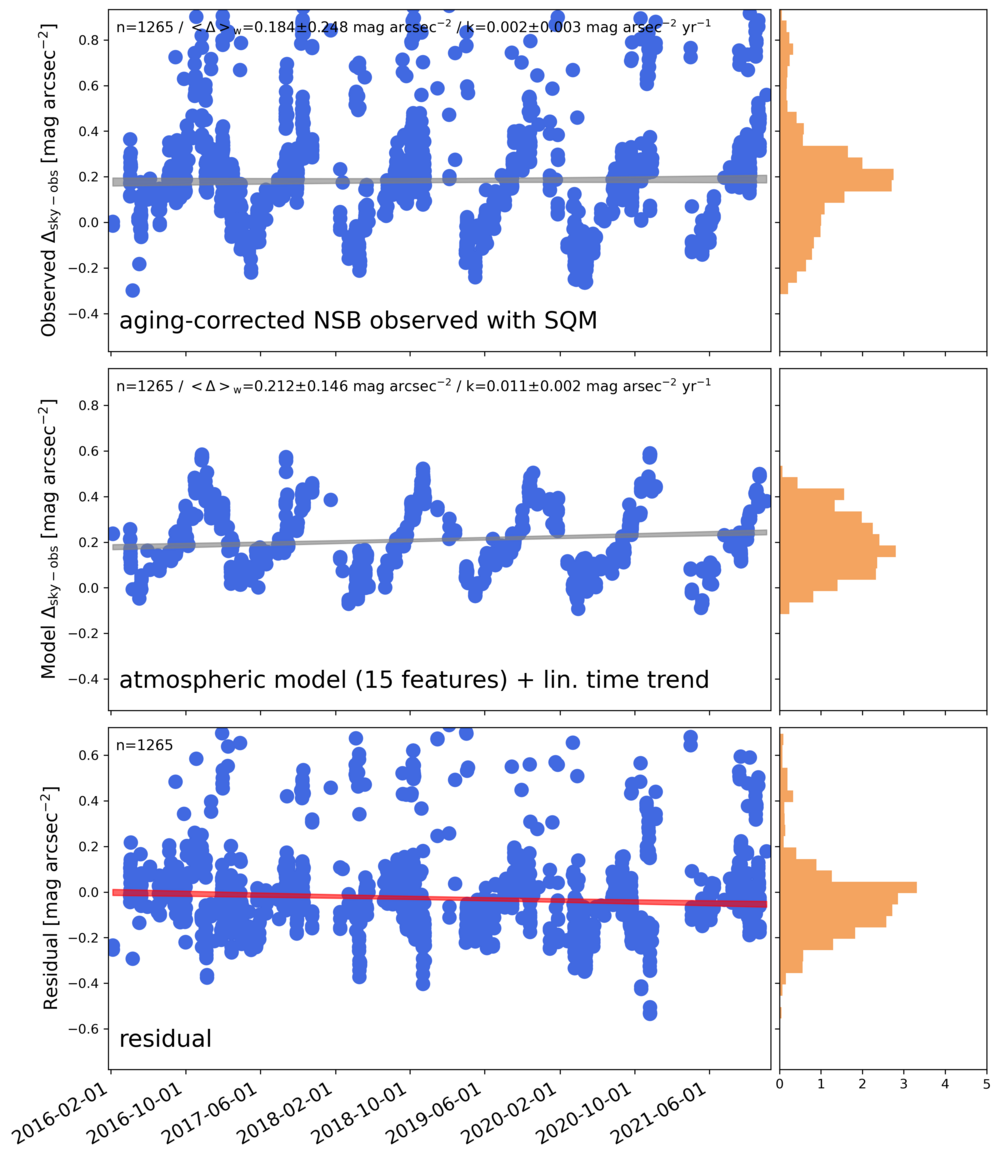}
        \caption[Long-term Trendanalysis]{Long-term trend for GRI and GRU. See caption Figure \ref{fig:BA1_BOD_trend} for more details.}
        \label{fig:GRI_GRU_trend}
\end{figure*}

\begin{figure*}
\centering
        \includegraphics[width=1\columnwidth]{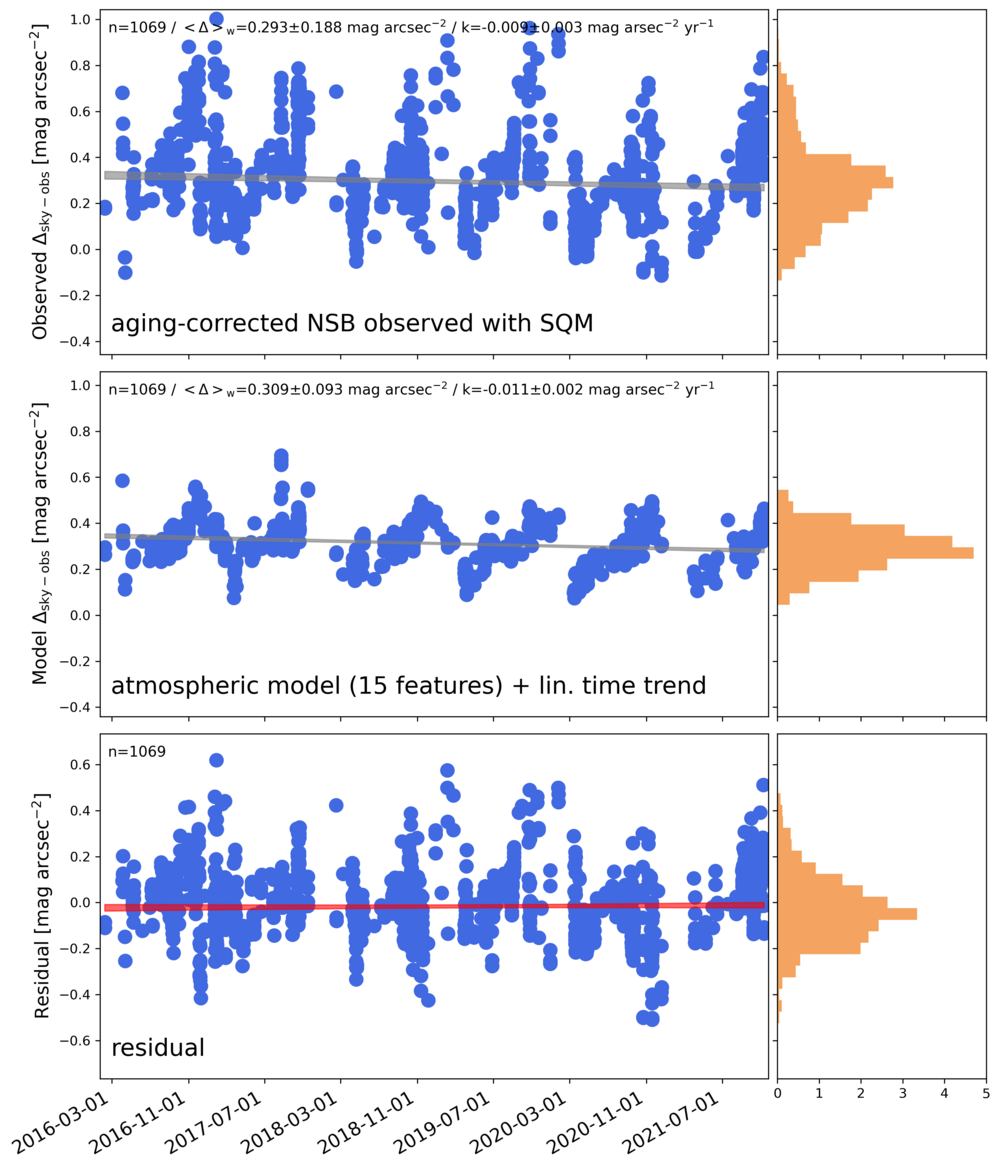}
        \includegraphics[width=1\columnwidth]{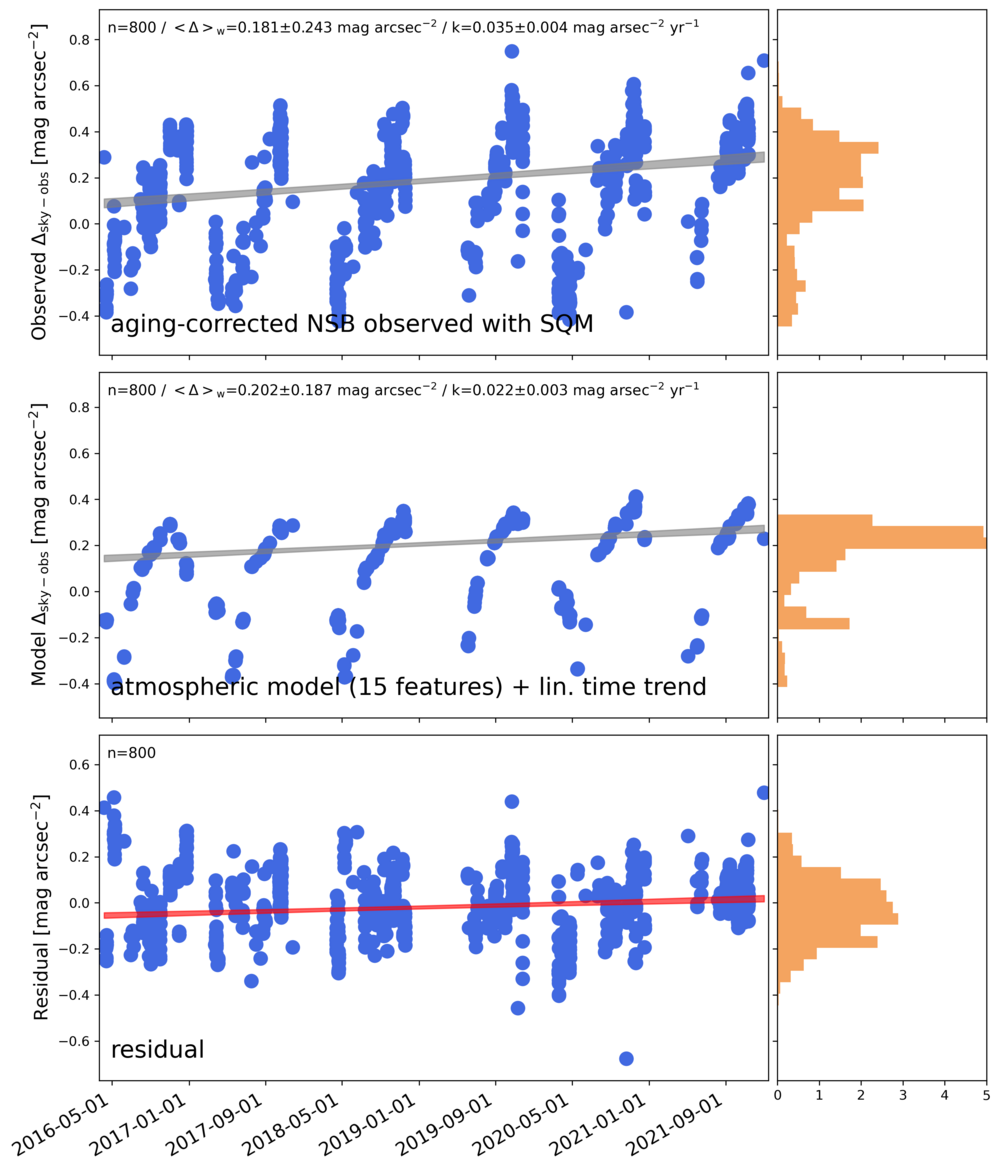}
        \caption[Long-term Trendanalysis]{Long-term trend for KID and KRI. See caption Figure \ref{fig:BA1_BOD_trend} for more details.}
        \label{fig:KID_KRI_trend}
\end{figure*}

\begin{figure*}
\centering
        \includegraphics[width=1\columnwidth]{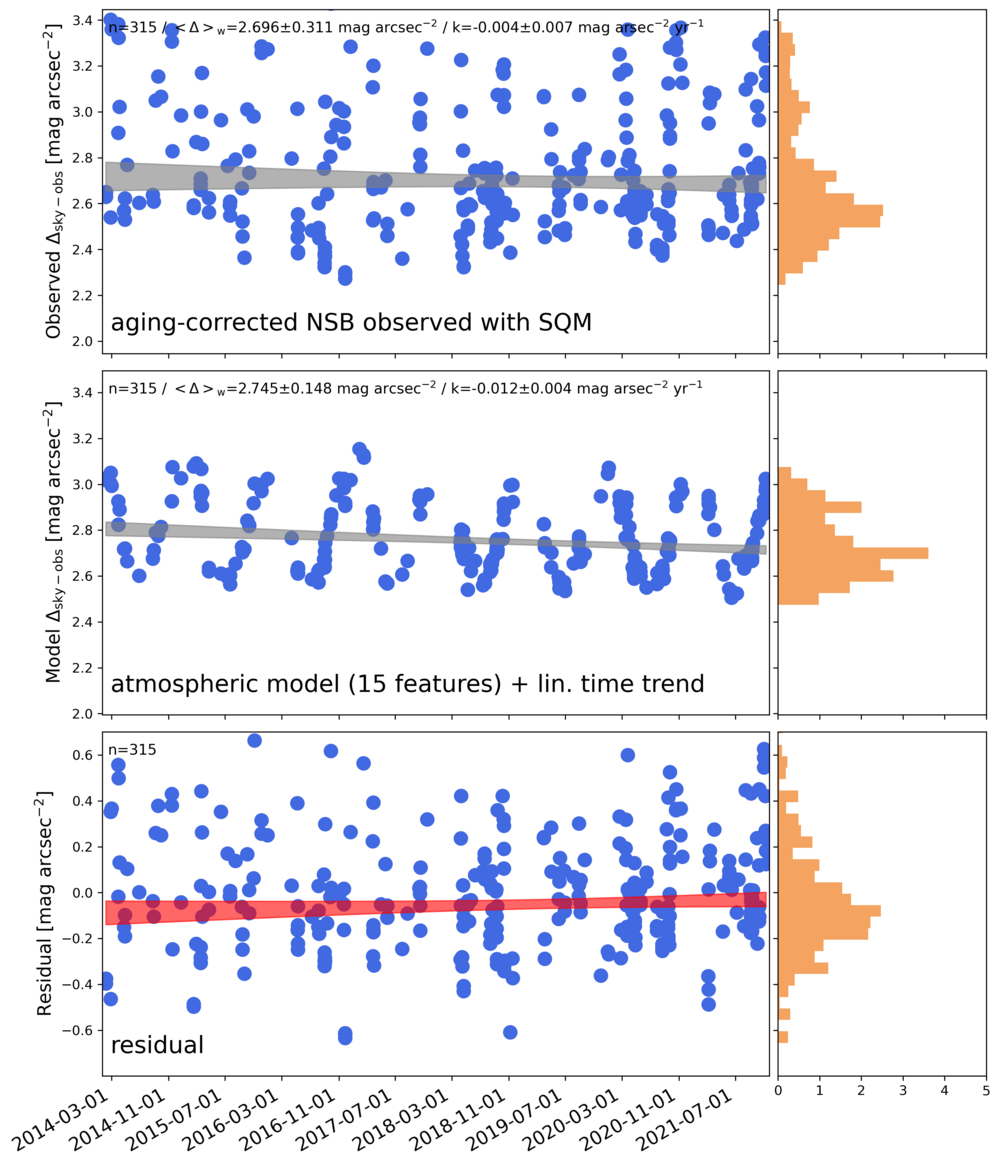}
        \includegraphics[width=1\columnwidth]{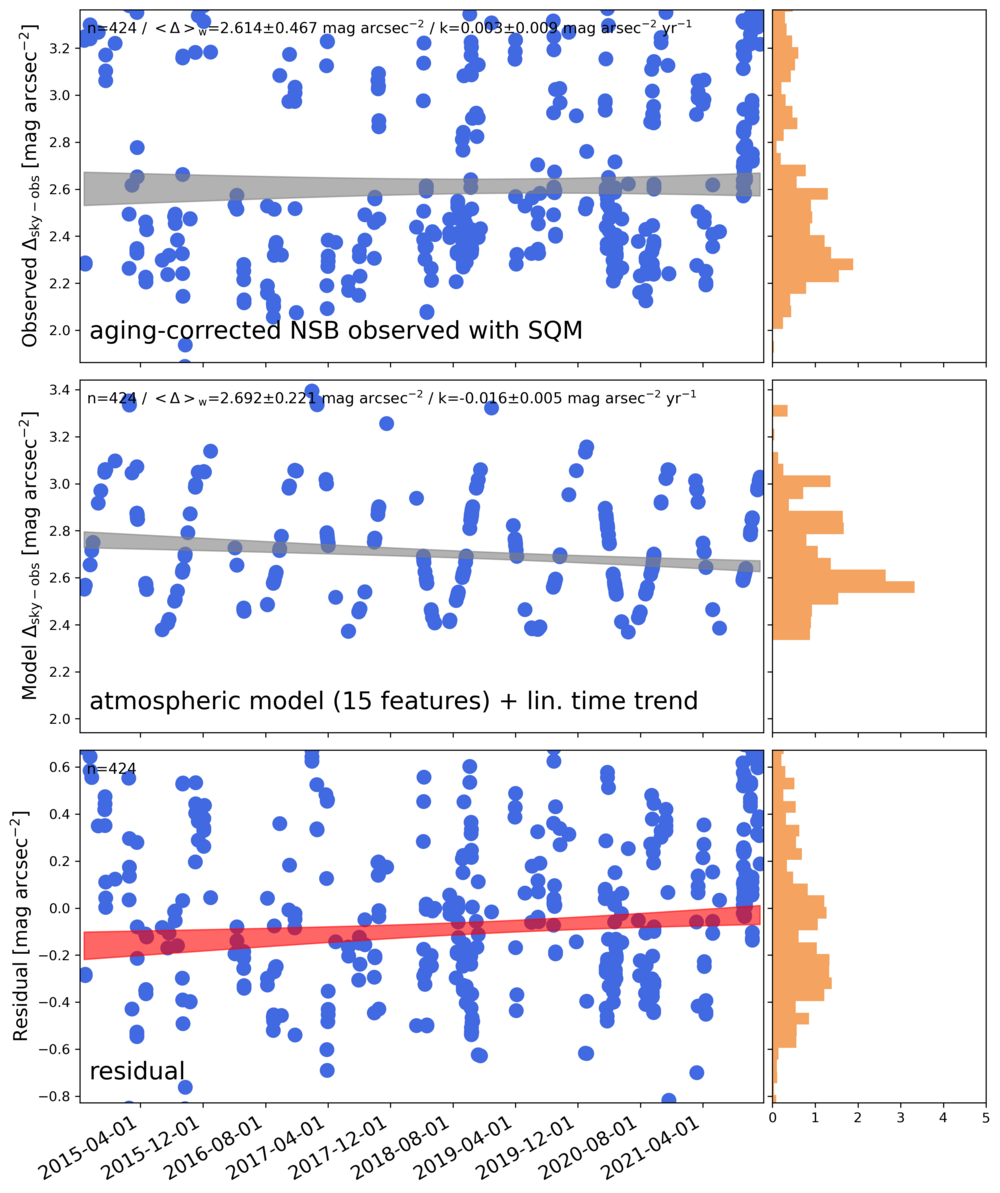}
        \caption[Long-term Trendanalysis]{Long-term trend for LGO and LSM. See caption Figure \ref{fig:BA1_BOD_trend} for more details.}
        \label{fig:LGO_LSM_trend}
\end{figure*}

\begin{figure*}
\centering
        \includegraphics[width=1\columnwidth]{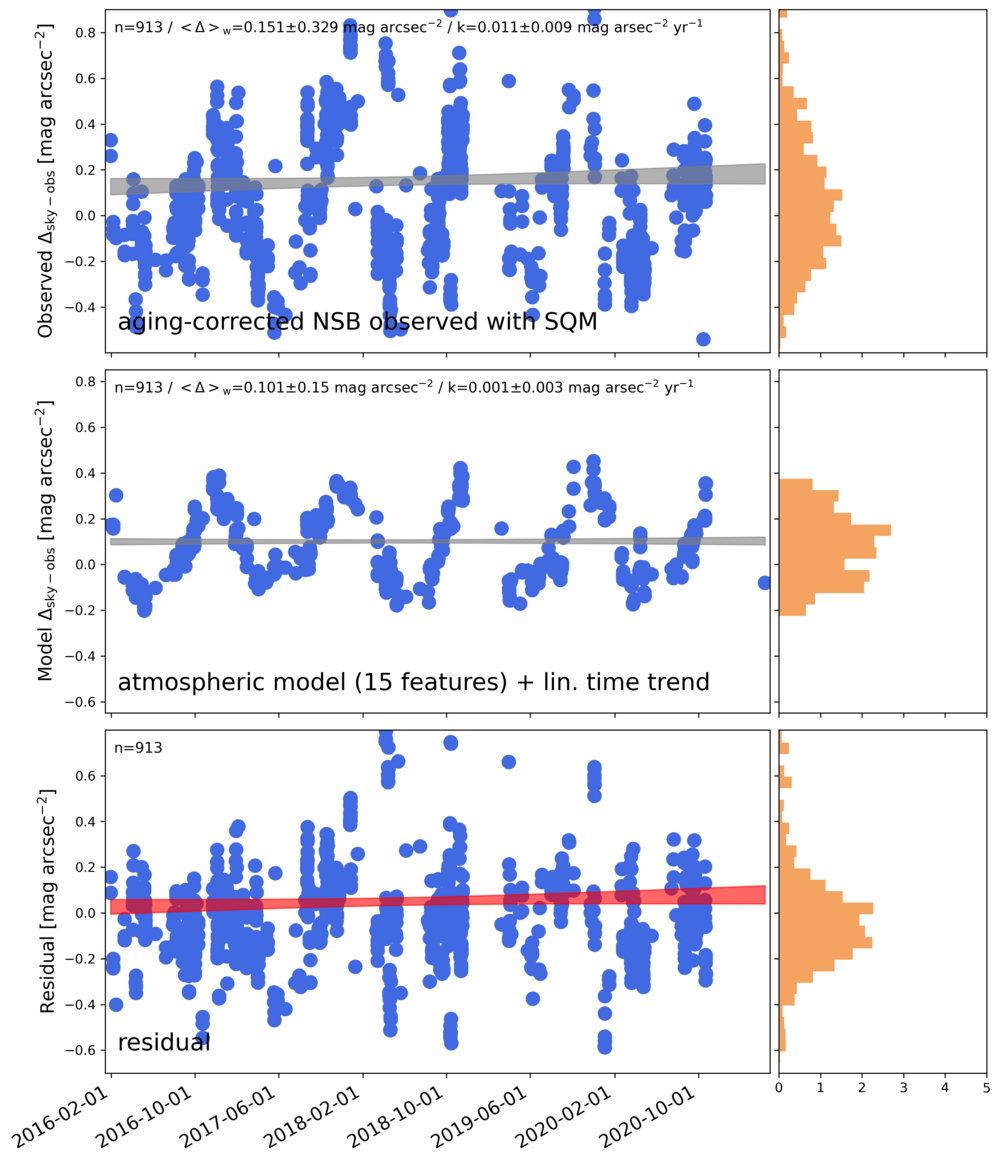}
        \includegraphics[width=1\columnwidth]{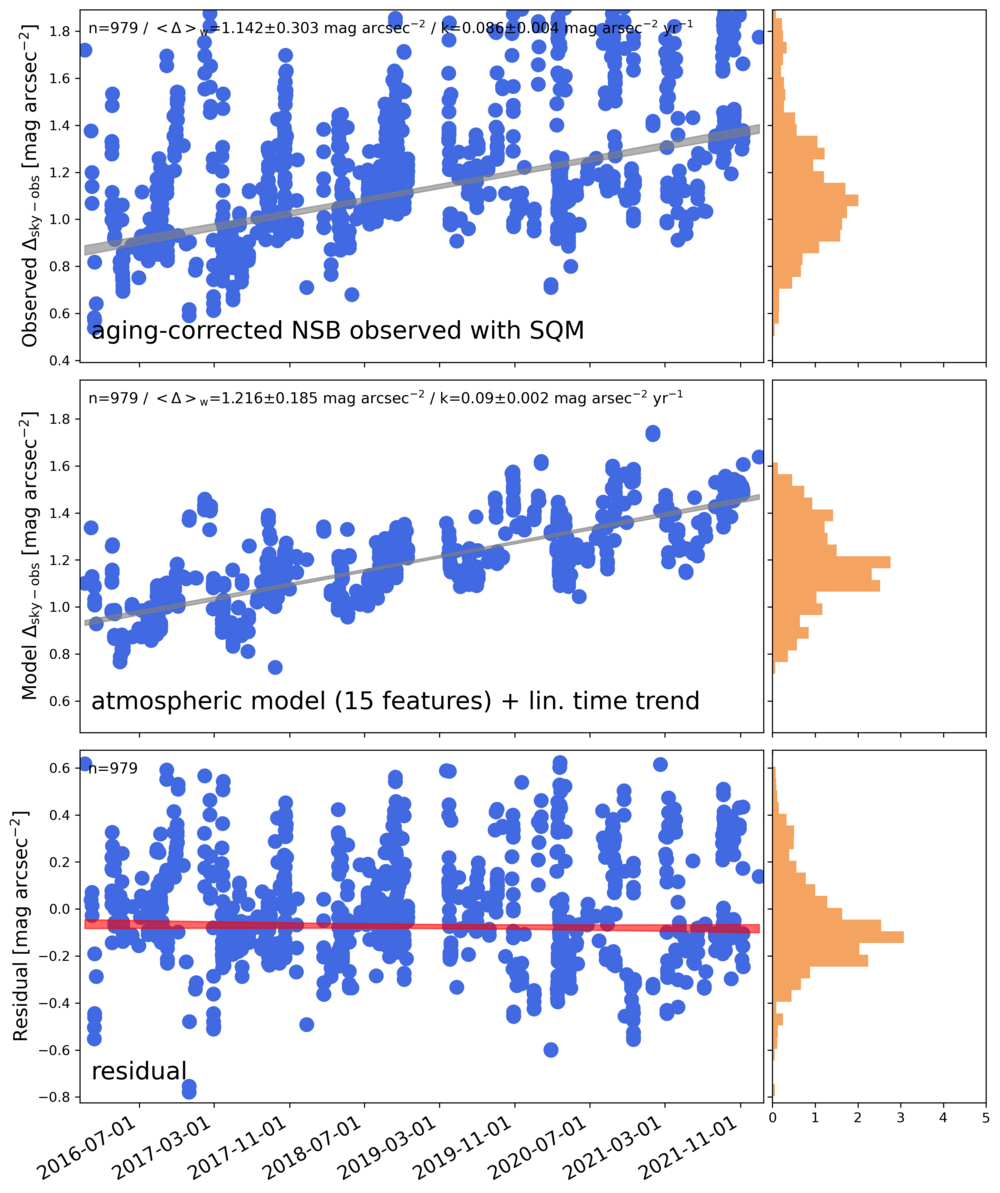}
        \caption[Long-term Trendanalysis]{Long-term trend for LOS and MAT. See caption Figure \ref{fig:BA1_BOD_trend} for more details.}
        \label{fig:LOS_MAT_trend}
\end{figure*}

\begin{figure*}
\centering
        \includegraphics[width=1\columnwidth]{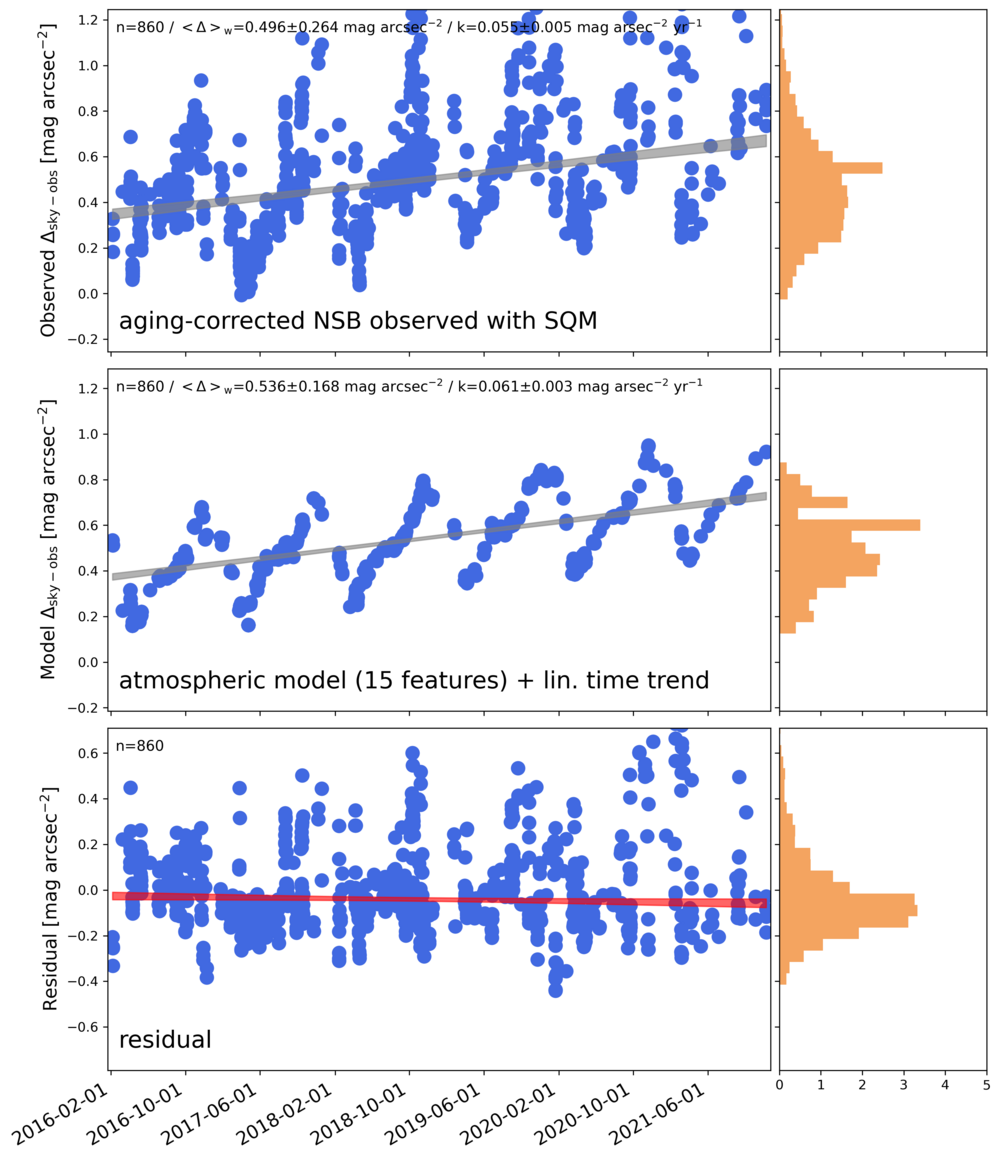}
        \includegraphics[width=1\columnwidth]{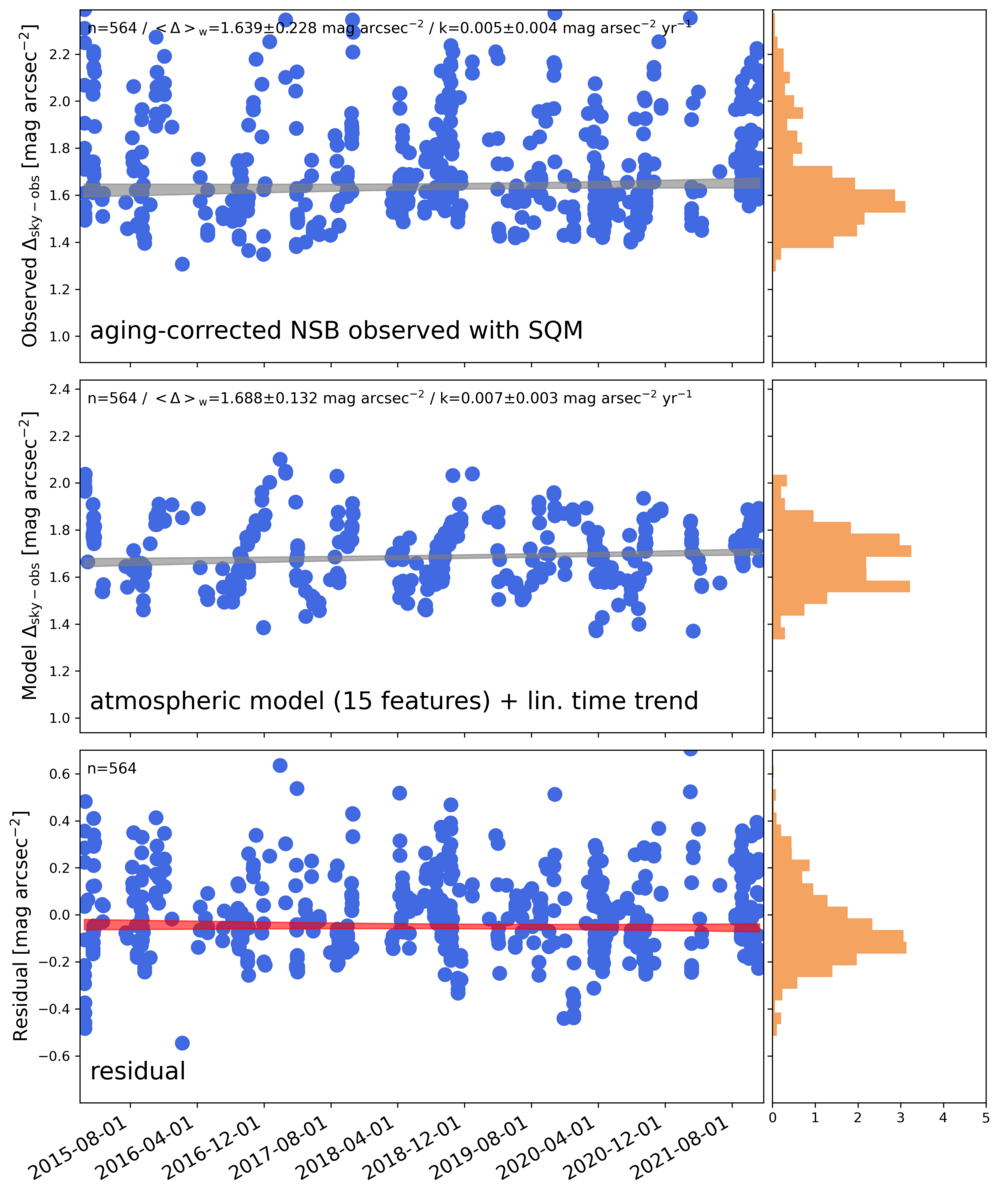}
        \caption[Long-term Trendanalysis]{Long-term trend for MUN and PAS. See caption Figure \ref{fig:BA1_BOD_trend} for more details.}
        \label{fig:MUN_PAS_trend}
\end{figure*}

\begin{figure*}
\centering
        \includegraphics[width=1\columnwidth]{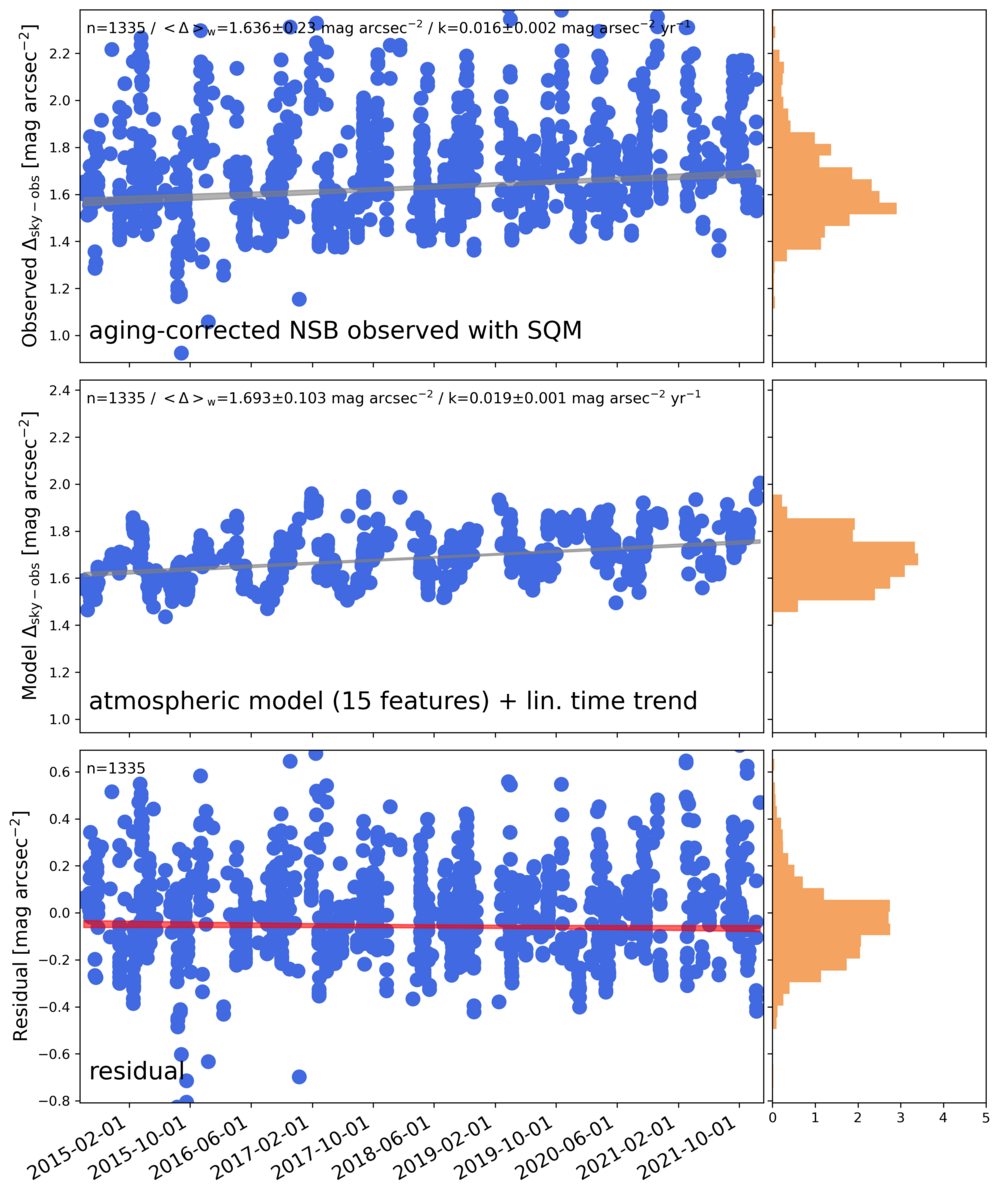}
        \includegraphics[width=1\columnwidth]{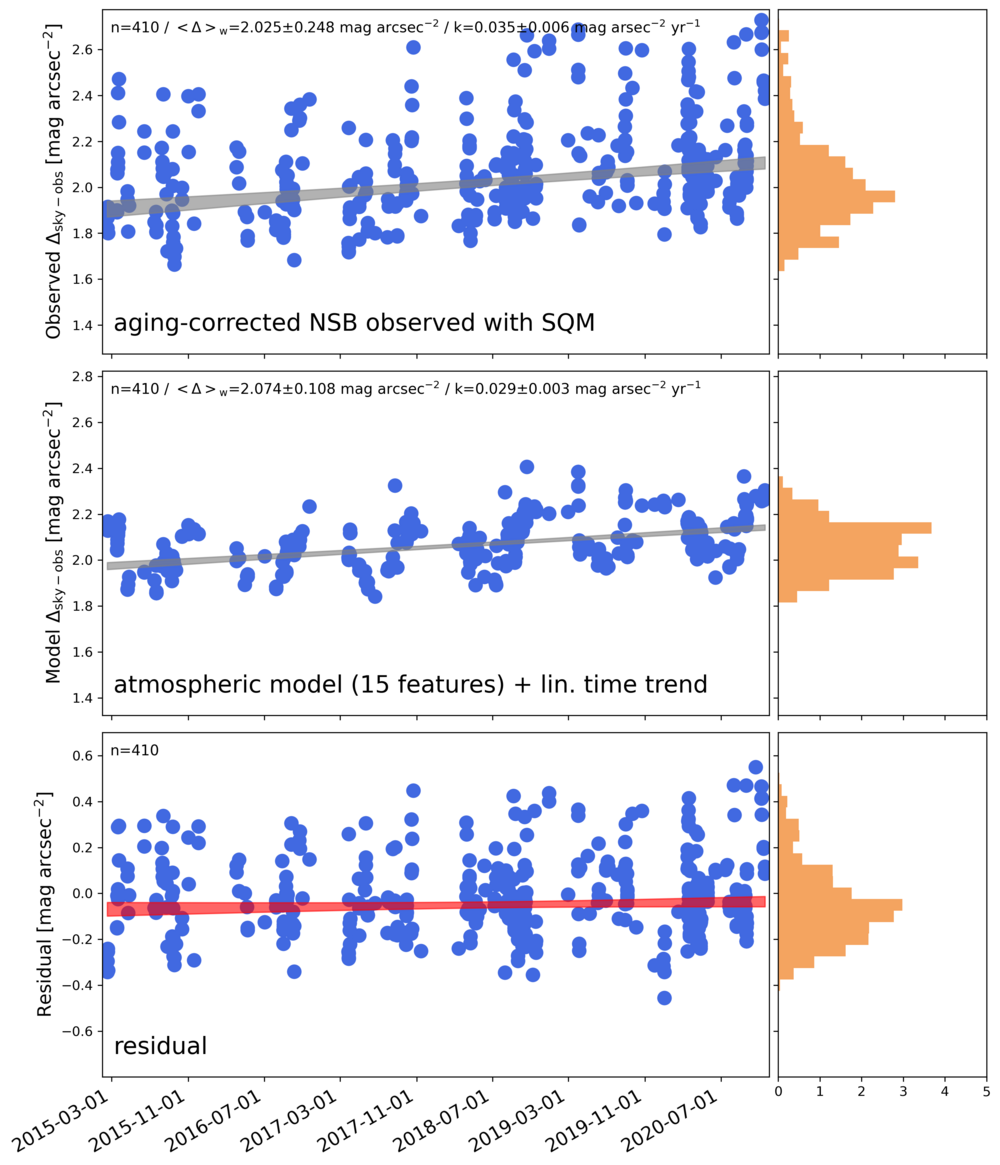}
        \caption[Long-term Trendanalysis]{Long-term trend for STY and TRA. See caption Figure \ref{fig:BA1_BOD_trend} for more details.}
        \label{fig:STY_TRA_trend}
\end{figure*}

\begin{figure*}
\centering
        \includegraphics[width=1\columnwidth]{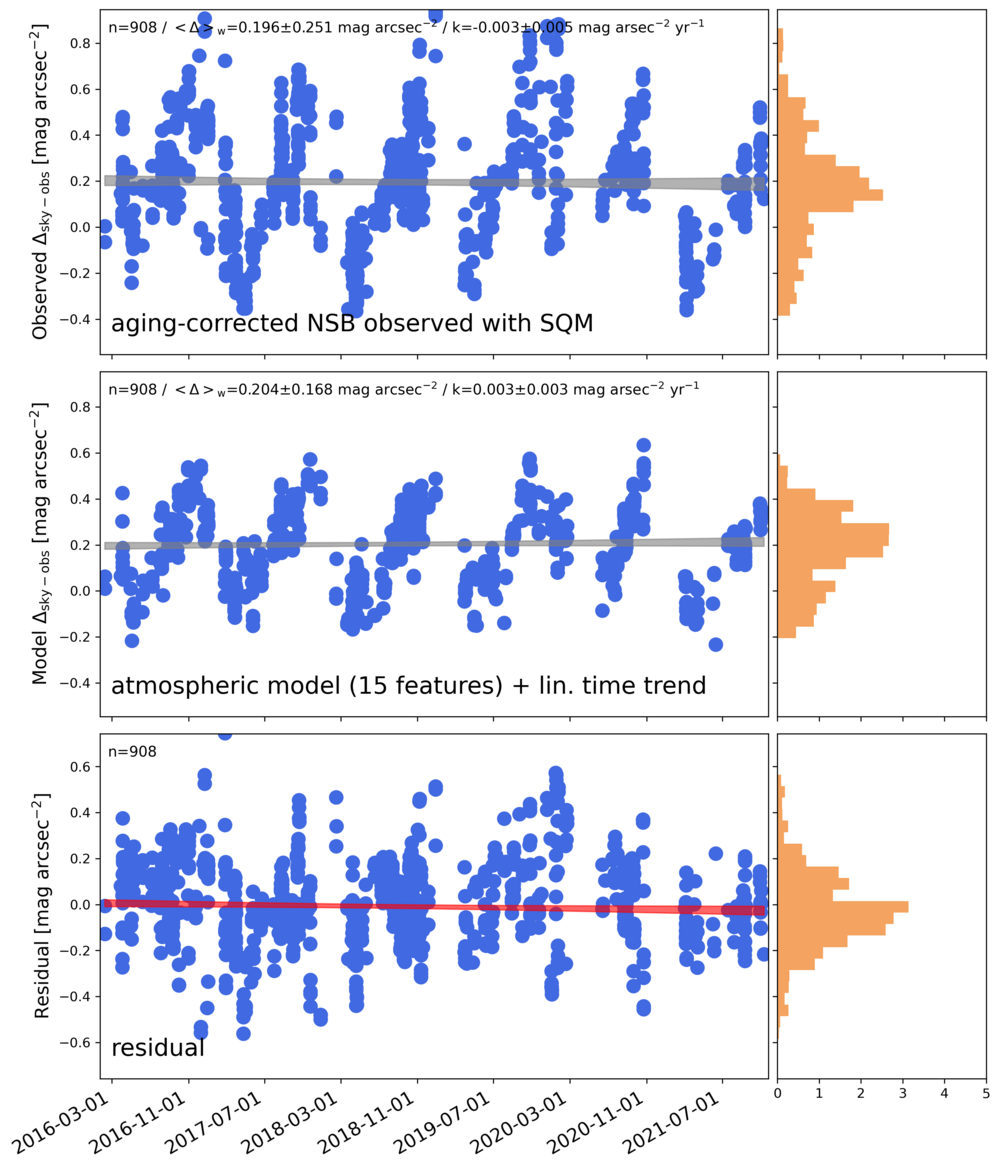}
        \includegraphics[width=1\columnwidth]{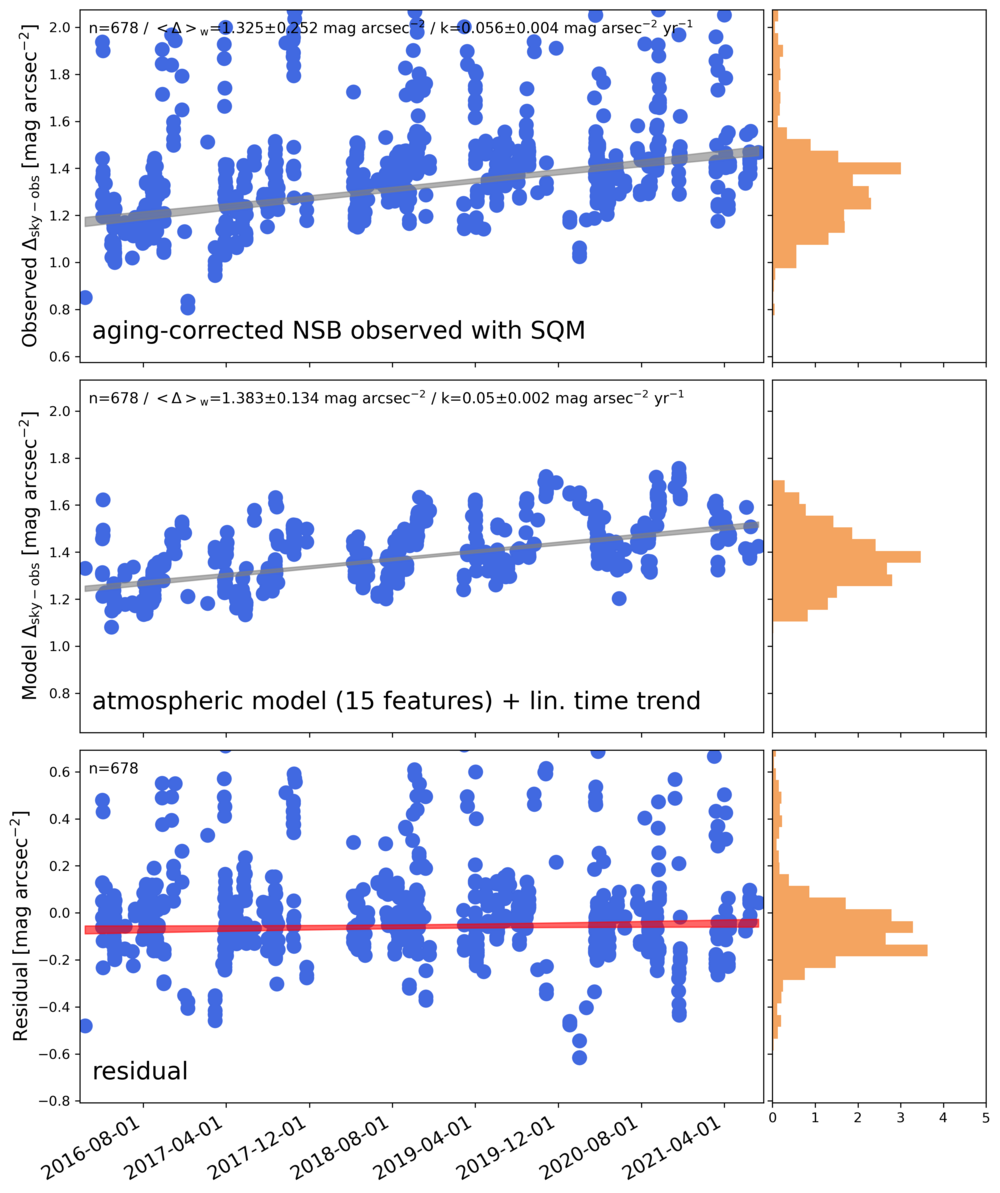}
        \caption[Long-term Trendanalysis]{Long-term trend for ULI and VOE. See caption Figure \ref{fig:BA1_BOD_trend} for more details.}
        \label{fig:ULI_VOE_trend}
\end{figure*}

\begin{figure*}
\centering
        \includegraphics[width=1\columnwidth]{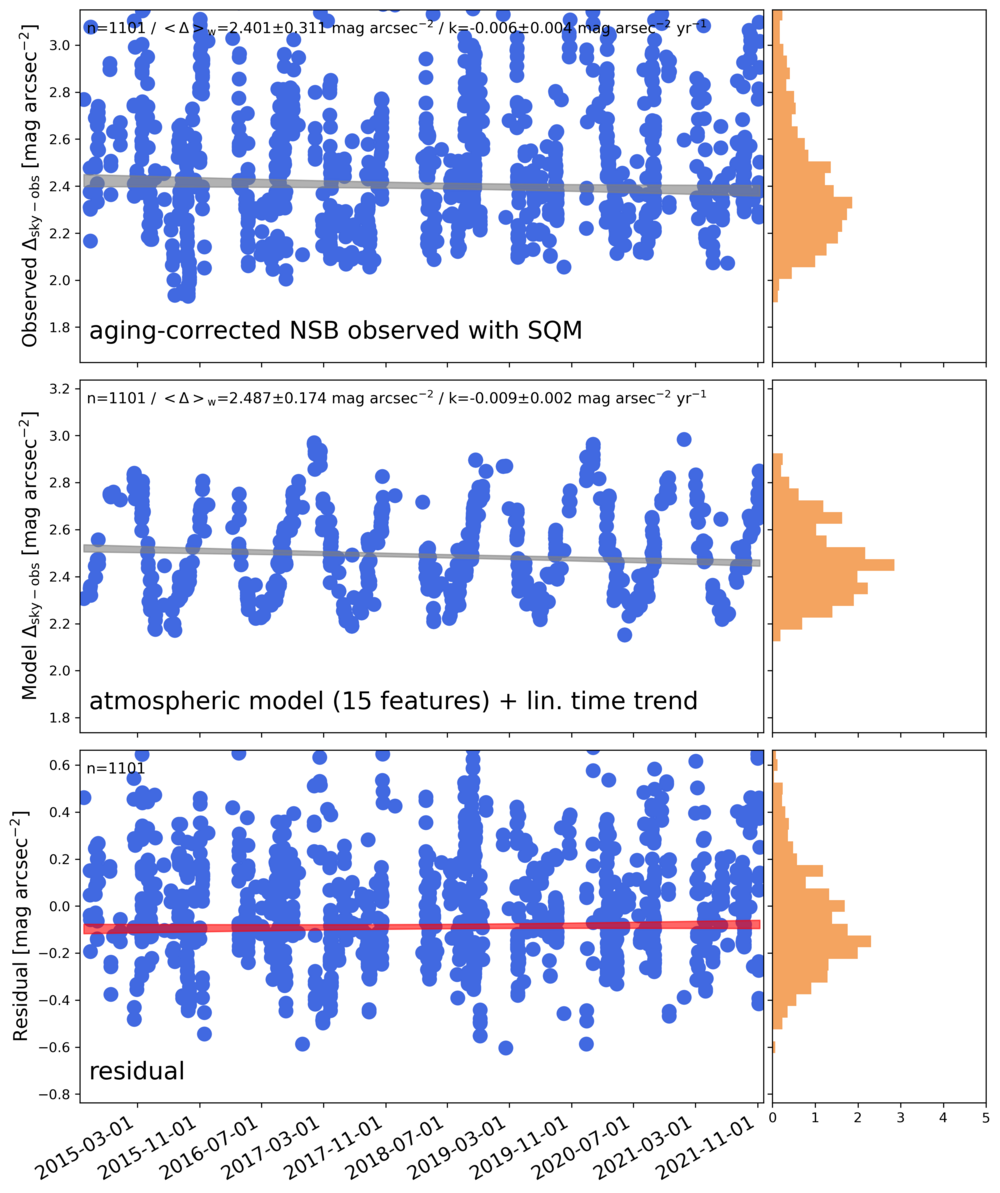}
        \includegraphics[width=1\columnwidth]{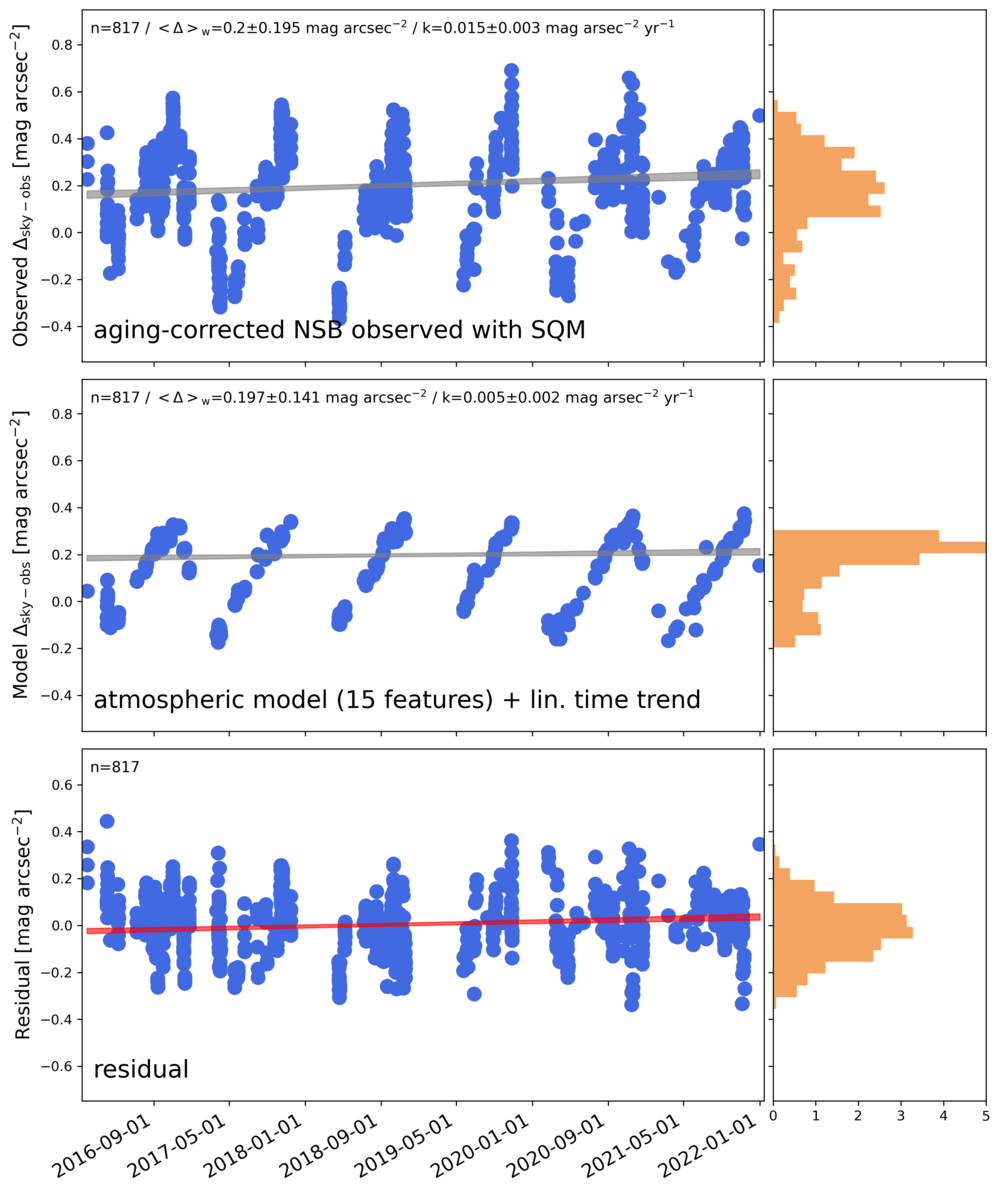}
        \caption[Long-term Trendanalysis]{Long-term trend for WEL and ZOE. See caption Figure \ref{fig:BA1_BOD_trend} for more details.}
        \label{fig:WEL_ZOE_trend}
\end{figure*}

\bsp	
\label{lastpage}
\end{document}